\documentclass{aa}  

\usepackage{graphicx}
\usepackage{color}
\usepackage{longtable}
\usepackage{supertabular}
\usepackage{natbib}
\usepackage[normalem]{ulem}

\definecolor{linkcolor}{rgb}{0.2, 0.55, 0.9}
\usepackage[breaklinks, colorlinks, citecolor=linkcolor]{hyperref}

\usepackage[varg]{txfonts}

\defcitealias{MBB13}{MBB} 
\defcitealias{Cava+17}{Paper~I}
\defcitealias{Mamon+19}{Paper~II}

\usepackage{txfonts}
\begin{document} 
\definecolor{red}{rgb}{1.,0.,0.}
\definecolor{darkred}{rgb}{0.9,0,0.3}
\definecolor{orange}{rgb}{0.9,0.4,0.0}
\definecolor{darkblue}{rgb}{0.,0.,0.3}
\definecolor{darkgreen}{rgb}{0,0.5,0}
\definecolor{grey}{rgb}{0.45,0.55,0.65}
\definecolor{Turquoise}{rgb}{0.2,0.9,0.8}
\definecolor{RoyalBlue}{rgb}{0.25,0.4,0.9}
\definecolor{NavyBlue}{rgb}{0.,0,0.6}
\definecolor{ForestGreen}{rgb}{0.,0.6,0.}
\newcommand{\mr}{M(r)}
\newcommand{\ms}{{\rm M}_{\odot}}
\newcommand{\br}{\beta(r)}
\newcommand{\bpr}{\sigma_{\rm r}/\sigma_{\theta}}
\newcommand{\rvir}{r_{200}}
\newcommand{\rtwo}{r_{-2}}
\newcommand{\cvir}{c_{200}}
\newcommand{\vvir}{v_{200}}
\newcommand{\mvir}{M_{200}}
\newcommand{\rnu}{r_{\nu}}
\newcommand{\rb}{r_{\beta}}
\newcommand{\rs}{r_{\rm s}}
\newcommand{\ks}{\mathrm{km \, s}^{-1}}
\newcommand{\slos}{\sigma_{\rm{los}}}
\newcommand{\vrf}{v_{{\rm rf}}}
\newcommand{\qr}{Q_r}
\newcommand{\qrho}{Q_{\rho}}
\newcommand{\qnu}{Q_{\nu}}
\newcommand{\qrrho}{Q_{r,\rho}}
\newcommand{\qrnu}{Q_{r,\nu}}
\newcommand{\gnu}{\gamma_{\nu}}
\newcommand{\grho}{\gamma_{\rho}}
\newcommand{\qt}{Q_{{\rm th}}}
\newcommand{\qtr}{Q_{{\rm r,th}}}
\newcommand{\ar}{\alpha_{\rm{r}}}
\newcommand{\abi}[1]{\textcolor{magenta}{#1}}
\newcommand{\ab}[1]{\textcolor{darkred}{\bf Andrea: #1}}
\newcommand{\gam}[1]{\textcolor{blue}{\bf Gary: #1}}
\newcommand{\gm}[1]{\textcolor{blue}{#1}}
\newcommand{\hide}[1]{\textcolor{grey}{#1}}
\newcommand{\old}[1]{\textcolor{red}{\sout{#1}}}
\newcommand{\new}[1]{\textcolor{darkgreen}{#1}}

   \title{Structural and dynamical modeling of WINGS clusters}

   \subtitle{III. The pseudo phase-space density profile}

   \author{A. Biviano
          \inst{1,2,3}
          \and
          G.~A.~Mamon \inst{3}
          }

   \institute{
   INAF-Osservatorio Astronomico di Trieste, via G. B. Tiepolo 11, 34143 Trieste, Italy\\ \email{andrea.biviano@inaf.it}
   \and
   IFPU-Institute for Fundamental Physics of the Universe, via Beirut 2, 34014 Trieste, Italy
   \and
   Institut d'Astrophysique de Paris (UMR 7095: CNRS \& Sorbonne Universit\'e), 98 bis Bd Arago, 75014 Paris, France
               }

   \date{Received \ldots; accepted \ldots}

\abstract{Numerical simulations indicate that cosmological halos display power-law radial profiles of pseudo phase-space density (PPSD),  $Q \equiv \rho/\sigma^3$, where $\rho$  is mass density and $\sigma$ velocity dispersion. We test these predictions for $Q(r)$ using the parameters derived from the Markov Chain Monte Carlo (MCMC) analysis performed with the MAMPOSSt mass-orbit modeling code on the observed kinematics of a velocity dispersion based stack ($\sigma_v$) of 54 nearby regular clusters of galaxies from the WINGS dataset. In the definition of PPSD, the density is either in total mass $\rho$ ($\qrho$) or in galaxy number density $\nu$ ($\qnu$) of three morphological classes of galaxies (ellipticals, lenticulars, and spirals), while the velocity dispersion (obtained by inversion of the Jeans equation using the MCMC parameters)  is either the total ($\qrho$ and $\qnu$) or its radial component ($\qrrho$ and $\qrnu$). We find that the PPSD profiles are indeed power-law relations for nearly all  MCMC parameters. The logarithmic slopes of our observed $\qrho(r)$ and $\qrrho(r)$ for ellipticals and spirals are in excellent agreement with the predictions for particles in simulations, but slightly shallower for S0s. For $\qnu(r)$ and $\qrnu(r)$, only the ellipticals have a PPSD slope matching that of particles in simulations, while the slope for spirals is much shallower, similar to that of subhalos. But for cluster stacks based on richness or gas temperature, the fraction of power-law PPSDs is lower (esp. $\qnu$) and the $\qrho$ slopes are shallower, except for S0s. The observed PPSD profiles, defined using $\rho$ rather than $\nu$, appear to be a fundamental property of galaxy clusters. They would be imprinted during an early phase of violent relaxation for dark matter and ellipticals, and later for spirals as they move towards dynamical equilibrium in the cluster gravitational potential, while S0s are either intermediate (richness and temperature-based stacks) or a mixed class ($\sigma_v$ stack).}


   \maketitle
%

\section{Introduction} \label{s:intro}
Cosmological dissipationless simulations have led to the building blocks of the standard model of dark matter, in particular through
the establishment of the universality of cosmic structure (halo) density profiles, well characterized by the NFW \citep*{NFW96} and  Einasto models \citep{Navarro+04}.\footnote{The Einasto model was first proposed by \cite{Einasto65} in a completely different context.} Further insight into the structure of clusters of galaxies have come from the analysis of \citet{TN01}, who examined the coarse-grained phase-space density profiles of cold dark matter (DM) halos from cosmological simulations. They defined the pseudo-phase space density (PPSD) profile 
$Q(r) \equiv \rho(r)/\sigma(r)^3$, where $\rho(r)$ and $\sigma(r)$ are the radial profiles of total mass density and velocity dispersion, respectively. They found $Q(r)$ to follow a power law $Q(r) \propto r^{\alpha}$ with $\alpha\approx-1.875$ over two and a half decades in radius. The equivalent PPSD built with the radial component of the velocity dispersion $\sigma_r$, i.e. $Q_r(r) \equiv \rho(r)/\sigma_r(r)^3$, is also found to obey a power-law relation with radial coordinate $r$,
with a  slightly steeper slope than $Q(r)$ \citep*{RTM04,DML05}.\footnote{For the sake of simplicity, in the rest of this introduction, unless otherwise specified, we use $Q$ to refer to both $Q$ and $\qr$.} These power-law behaviors are remarkable given that the logarithmic density profile $\log\,\rho(r)$ and the logarithmic velocity dispersion profiles $\log\,\sigma(r)$ (total) and $\log\,\sigma_r(r)$ (radial) are all convex functions of $\log\,r$. The  slope of $Q(r)$ matches the slope of $-15/8$ expected from  secondary infall models  based on (quasi)-power law density profile $\rho \sim r^{-9/4}$
\citep{Gott75,Gunn77,Bertschinger85}, despite the fact that DM halos in the cosmological simulations of \citet{TN01} assemble in a different way from the regular phase-space stratification process described by \citeauthor{Gott75}, \citeauthor{Gunn77}, and \citeauthor{Bertschinger85}.

Much effort has been employed in trying to understand why $Q(r)$ is a simple power-law, and why its slope is so close to the value predicted by the secondary infall model of \citet{Bertschinger85}. Analytical and numerical studies have shown that the final shape of $Q(r)$ does not depend on whether the halo evolves through major mergers or spherical infall \citep{Manrique+03,AYGM04,Austin+05,HRDSH07}. The final $Q(r)$ configuration of cosmological halos appears to be reached very early on, soon or immediately after the early relaxation phase \citep{VVKK09,Colombi21}, driven by violent relaxation \citep{LyndenBell67}. Other collective relaxation processes might be important as well in shaping $Q(r)$, such as Landau damping and radial orbit instability \citep*{Henriksen06,MMWH06}. At large radii, where relaxation is still incomplete, 
deviation from the power-law behavior was expected theoretically \citep{Bertschinger85,LC11}, and also detected in numerical simulations. However, deviations from the power-law behavior never exceed 20\% within the halo virial radius \citep{Ascasibar&Gottlober08,Navarro+10,Ludlow+10,Marini+21}. Close to the halo center, where baryonic effects can be important, both a steepening \citep{LC11} and a flattening \citep{Butski+16} of $Q(r)$ have been predicted.

\citet{VVKK09} suggested that the original physical association of $Q(r)$ with the halo coarse phase-space density is not justified, as the two quantities have different behaviors \citep[but see][]{MH09}. Interpretation of $Q(r)$ in terms of the power $-3/2$ of the dynamical entropy of a gravitating system might prove more useful to understand its origin. \citet{FHGY07} noted the similarity in the so-defined dynamical entropy of DM particles and the thermodynamic entropy of the intra-cluster gas, outside the core of simulated clusters. 
\citet{HK10} argued that $Q(r)$ corresponds to a minimum entropy state, that is the end result of
long-range (e.g. violent) relaxation processes in gravitating systems, while the state of maximum entropy is only reached locally, where short-range (e.g. two-body) relaxations dominate.

The analysis of simulations leads to contradicting conclusions on the universality of $Q(r)$ slopes. \citet*{Schmidt+08} argued that $Q(r)$ is not universal, while \citet{DML05,Navarro+10,Marini+21} found very similar slopes for the $Q(r)$ of different halos (with a difference of $\lesssim 15$~\%).
The $Q(r)$ slope is only mildly dependent ($\approx \pm 10$\% ) on redshift \citep{LC09,Marini+21} and on the power spectrum of primordial density fluctuations \citep{KKH08,Brown+20}.

Almost all numerical investigations so far have been focused on the $Q(r)$ traced by DM particles,
and similar results for $Q(r)$ have been obtained in DM-only and in hydrodynamical simulations \citep[compare, e.g.,][]{DML05,RTM04}. Only recently, different tracers have been considered in the definition of $Q(r)$ in the
study of \citet{Marini+21}, and the $Q(r)$ slope has been found to be
strongly dependent on the chosen tracer, being steeper for stars and shallower for galaxies (subhalos in hydrodynamical simulations), than for DM particles
($\alpha=-3, -1.3,$ and $-1.8$, respectively).
 This dependence is very relevant when comparing simulation-based predictions with observations, since $Q(r)$ is not an observable; $\rho(r)$ can be inferred from stellar and galaxy  kinematics, from gravitational lensing, or from the temperature and pressure of the intra-cluster gas \citep[see, e.g.][for a review]{Pratt+19}, but $\sigma(r)$ can only be determined for the tracer of the gravitational potential. Since only the tracer $\sigma(r)$ can be determined observationally, for consistency 
some authors have used the number density profile of the tracers, $\nu(r)$, rather than the total mass density profile $\rho(r)$, in the definition of  $Q(r)$.

Several observational studies have confirmed the expected simulation-based power-law behavior of $Q(r)$, both for galaxies and for clusters of galaxies. \citet{Chae14}
found that $Q(r)$ is a power law for Coma cluster elliptical galaxies with a slope of $1.93\pm0.06$. 
On larger scales, $Q(r)$ has been measured in several clusters of galaxies over the redshift range 0.06--1.34, and always found to be similar to, or at least consistent with, the simulation-based expectations, both when $Q(r)$ was defined using the total mass density profile $\rho(r)$ (\citealt{Biviano+13}; \citealt*{MBM14}; \citealt{Biviano+16,Biviano+21}), and when the tracer $\nu(r)$ was used instead, for tracers of different colors and luminosities (\citealt*{MBM15}; \citealt{AADDV17,Capasso+19}).

Despite the good agreement of the simulation-based predictions with observations, the power-law behavior, and even the physical reality of $Q(r)$, have been questioned. \citet*{NOJ17} argued against a power-law behavior of $Q(r)$ at any radial scale, and argued that the agreement between $Q(r)$ found in numerical simulations and the solution of the secondary infall model of \citet{Bertschinger85} is purely coincidental. According to \citet{Schmidt+08} different halos follow $\rho/\sigma_r^{\epsilon} \propto r^{\alpha}$ relations, with different best-fit values of $\epsilon$ and $\alpha$, that is, $\epsilon=3$ is not a universal value. \citet{AW20} argued that the power-law behavior of $Q(r)$ does not have a physical origin, but it is just a fluke, a consequence of the linear relation between the logarithmic slope of the mass density profile, $\gamma \equiv d \ln \rho / d \ln r$, and the velocity anisotropy profile 
$\beta = 1-\sigma_{\theta}^2/\sigma_r^2$, where $\sigma_{\theta}$ and $\sigma_r$ are the tangential and radial component of the velocity dispersion tensor. The linear $\beta-\gamma$ relation was discovered by \citet{HM06} and \citet{HS06} in a variety of simulated gravitating systems, issued from controlled simulations of halo-halo collisions and radial infall, as well as from cosmological simulations. However, the physical origin of the linear $\beta-\gamma$ relation is not better elucidated than that of the $Q(r)$ power-law, and the relation is not clearly established in real clusters \citep{Biviano+13,MBM14,AADDV17,Biviano+21}.

Lacking a clear understanding of the physical origin(s) of either $Q(r) \propto r^{\alpha}$
or the linear $\beta-\gamma$ relation, several studies have tried to investigate their consistency in the context of the dynamical equilibrium of a spherical gravitating system, as described by the Jeans equation, which for a spherical stationary system is \citep{Binney80}
\begin{equation}
    {{\rm d}\left(\nu \sigma_r^2\right)\over {\rm d}r} + 2\,\beta(r)\,{\nu\,\sigma_r^2\over r} = - \nu(r) {G\,M(r)\over r^2}.
\end{equation}
By assuming a linear $\beta-\gamma$ relation, \citet{DML05} found a critical solution that satisfies $\rho/\sigma_r^{\epsilon} \propto r^{\alpha}$, with the value of $\alpha$ being dependent only on $\epsilon$ and $\beta_0$, and independent of the slope of the $\beta-\gamma$ relation. In particular, $\epsilon=3$ and $\beta_0=0$ lead to $\alpha=1.94$, essentially the same value found in numerical simulations. \citet{Barnes+07} considered  density profiles of the NFW or Einasto form, but
could not find consistent solutions of the Jeans equation with a power-law $Q(r)$ and a linear $\beta-\gamma$ relation similar to the relations found in simulations. They suggested that the $\beta-\gamma$ relation for any single halo is not strictly linear, and that the $\beta-\gamma$ relation is not just a manifestation of a scale-free $Q(r)$.  \citet*{ZHS08} started from the power-law behavior of $Q(r)$ to show that a linear $\beta-\gamma$ relation is inconsistent with generalized NFW  density profiles \citep{Zhao96}, but consistent with Einasto profiles 
of index $n=6$ \citep[see, e.g., eq.(16) of][Paper II hereafter]{Mamon+19}. 

The behavior of $Q(r)$ should depend on the choice of tracer used to measure $\sigma(r)$ and $\sigma_r(r)$, and possibly the density profile, when the number density profile, $\nu(r)$, is used in the definition of $Q(r)$. But the influence of tracer choice on $Q(r)$ has not yet been addressed.

In this article, we investigate $Q(r)$ and $\qr(r)$ in 54 nearby clusters of galaxies ($0.04<z<0.07$) from the WINGS data set \citep{Fasano+06}, which \citet[][hereafter Paper I]{Cava+17} found to be regular systems. In a forthcoming article (Mamon \& Biviano, in prep.), we will investigate the $\beta-\gamma$ relation in a similar fashion.
We exploit the results of the kinematic analysis of \citetalias{Mamon+19} that determined the mass density and velocity anisotropy profiles, $\rho(r)$ and $\beta(r)$, of stack samples of these clusters, as well as the number density profiles for each of three morphological classes of galaxies, from Gaussian priors obtained from previous  fits in \citetalias{Cava+17} of model plus constant background of the photometric data for the same stacked clusters. For the first time, we determine $Q(r)$ and $\qr(r)$ separately for three different morphological classes of cluster galaxies.

In the rest of this paper we use $Q$ and $\qr$ to refer generically to the pseudo-phase-space density profiles without distinction to whether they have been derived using $\rho(r)$ or $\nu(r)$. When needed, we will use subscripts to distinguish the different profiles, $\qrho, \qrrho$ and $\qnu, \qrnu$. 

The structure of this paper is the following. In Sect.~\ref{s:dat} we describe our data set, in Sect.~\ref{s:ana} our method of analysis. In Sect.~\ref{s:res} we present our results. We discuss our results in Sect.~\ref{s:dis}. Sect.~\ref{s:conc} provides a summary and our conclusions.

Throughout this paper we adopt the following
  cosmological parameters: $\Omega_{\rm m}=0.3, \Omega_{\Lambda}=0.7, H_0 =
  70\,\rm km\,s^{-1}\,Mpc^{-1}$. 

\section{The data set} \label{s:dat}
Our analysis is based on the WINGS data set, which contains X-ray-selected clusters at $0.04<z<0.07$ \citep{Fasano+06} with spectroscopic coverage for cluster galaxies with a median stellar mass of $\log( M_{\star}/\ms)=10.0$ for ellipticals (E) and 10.4 for spirals (S) \citepalias[\citealt{Cava+09,Moretti+14};][]{Mamon+19}. Morphological types were derived by \citet{Fasano+06} using the MORPHOT tool. In \citetalias{Cava+17}, we defined three intervals in the MORPHOT classification parameter corresponding to the three morphological classes of ellipticals, lenticulars (S0), and spirals.

In \citetalias{Cava+17}, we identified cluster members using the Clean algorithm \citep*{MBB13} and selected a subsample of 68 clusters with at least 30 spectroscopic members. Using the substructure test of \citet{DS88}, we identified 54 regular and 14 irregular clusters. We then estimated $\rvir$ and $\vvir$ of these 68 clusters in three different ways: based on (i) the cluster velocity dispersion (\texttt{sigv}), (ii) an estimate of the cluster richness (\texttt{num}, Mamon et al. in prep., see \citealt{Old+14}), and (iii) the cluster X-ray temperature (\texttt{tempX}, only available for 38 of these clusters). Using these three estimates of $\rvir, \vvir$ in \citetalias{Cava+17}, we then stacked the 54 (38, in the case of \texttt{tempX} scaling) regular clusters
into three pseudo-clusters, by rescaling the projected radii and rest-frame velocities of cluster galaxies by their cluster $\rvir$ and $\vvir$, respectively. These three pseudo-clusters formed the data set for the kinematic modeling that we performed in \citetalias{Mamon+19}, using the MAMPOSSt algorithm of \citet*{MBB13}. Irregular clusters were not considered because MAMPOSSt is based on the Jeans equation, which being derived from the collisionless Boltzmann equation, assumes that the tracers are  test particles orbiting the gravitational potential and not interacting with one another in contrast to galaxies within a substructure of an irregular cluster.

MAMPOSSt fits parametric models to the distributions of galaxies in projected phase space (projected distance to the center and line of sight velocity). The parameters are those describing the total mass density profile, $\rho(r)$, the tracer density profiles for the three morphological types ($i$), $\nu_i(r)$, and the velocity anisotropy profiles for the three types, $\beta_i(r)$. MAMPOSSt speeds up the calculations by a large factor by assuming that the three spherical-coordinate components of the local velocity distribution function are Gaussian. The recovered radial profiles of mass density and velocity anisotropy are as good with MAMPOSSt as with other methods \citep{Read+21}, even though MAMPOSSt is much faster.

Here we use the results of the kinematic modeling of \citetalias{Mamon+19}. More specifically, we consider the $M(r)$ and $\beta(r)$ model parameters of the outputs (chain elements) of the Markov Chain Monte Carlo  (MCMC) investigation of parameter space used by MAMPOSSt, using CosmoMC \citep{Lewis+02}, based on the Metropolis-Hastings algorithm.
This allows us to determine $Q(r)$ and $Q_r(r)$ at several values of $r$, and for the three different morphological classes, as detailed in Sect.~\ref{s:ana}.  For $Q_\nu$ and $\qrnu$, we also used the tracer number density profiles, $\nu_i(r)$, for each morphological class, obtained from  fitting NFW models plus constant background to the photometric data and then refined in the MCMC analysis with MAMPOSSt.

For each model, MAMPOSSt was run using 6 MCMC parallel chains, each with over $10^5$ elements, for a total of over 500\,000 chain elements (i.e. points in parameter space) per model.
We discard the 20\% of the first elements of each chain of each model, which corresponds to the `burn-in' phase where the MCMC has not yet settled to its equilibrium  and estimate $Q(r)$ and $Q_r(r)$ for all remaining chains.

We consider all three stacks obtained using the three scalings, \texttt{sigv, num, tempX}. 
We present the results for the \texttt{sigv} scaling in the main text
and discuss them in Sect.~\ref{ss:sigv}. Results for the
\texttt{num} and \texttt{tempX} scalings are presented in Appendix~\ref{s:scalings} and discussed in Sect.~\ref{ss:numtempx} in comparison with the results obtained for the \texttt{sigv} scaling.

\section{Analysis} \label{s:ana}
The parameter values in each MCMC are used to directly derive the radial profile $\nu(r), \rho(r)$, and $\beta(r)$. To determine $Q(r)$ and $Q_r(r)$, we also derive $\sigma(r)$ and $\sigma_r(r)$ via:
\begin{equation}
\sigma_r^2(r) = \frac{1}{\nu(r)} \, \int_r^\infty \exp \left[ 2 \, \int_r^s \beta(t) \frac{{\rm d} t}{t} \right] \, \nu(s) \frac{G M(s)}{s^2} \, {\rm d} s \ ,
\label{e:sigmar}
\end{equation}
\citep{vanderMarel94,ML05b}
and
\begin{equation}
    \sigma^2(r) = \left[3- 2 \, \beta(r)
    \right]\, \sigma_r^2(r) \ ,
    \label{e:sig}
\end{equation}
where $M(r)$ is the total mass profile.

\citetalias{Mamon+19} considered 30 sets of priors 
according to chosen models for the mass density profile:
\begin{equation}
     \rho(r) \propto r^{\gamma_0} \left(r+r_{\rm s} \right)^{\gamma_{\infty}-\gamma_0} \ ,
    \label{rhoofrmodel}
\end{equation}
with mass density logarithmic slope
\begin{equation}
    \gamma(r) = {\gamma_0 + \gamma_\infty\,x \over 1+x} \ ,
    \label{gNFWslope}
\end{equation}
where $x = r/r_s$, while $\gamma_0$ and $\gamma_\infty$ are the logarithmic slopes of the density profile at $r=0$ and at infinity, respectively.
The models considered in \citetalias{Mamon+19} nearly all assumed
$\gamma_{\infty}=-3$, $\gamma_0=-1$ for NFW and $\gamma_0 \neq -1$ for generalized NFW (`gNFW'), scale radius $r_s$ related to the radius where  $\gamma = -2$: $r_{-2} = (2+\gamma_0)\,r_s$.

We had also  adopted \cite{Einasto65}  mass models, which fit even better the distribution of radii in halos in $\Lambda$CDM dissipationless cosmological simulations \citep{Navarro+04},
\begin{equation}
   \rho(r) \propto \exp\left[-2\,n\,\left({r\over \rtwo}\right)^{1/n}\right]
   \, ,
\end{equation}
yielding
\begin{equation}
    \gamma(r) = -2 \left({r \over r_{-2}}\right)^{1/n} \ .
\end{equation}
The mass density models are normalized by the mass at radius $r_{200} = c_{200}\,r_{-2}$ where the mean mass density is equal to 200 times the critical density of the Universe at $z=0.055$, the median redshift of the WINGS clusters.

The anisotropy models considered in \citetalias{Mamon+19} followed
\begin{equation}
\beta(r) = \beta_0 + \left(\beta_\infty-\beta_0\right)\,{r^\delta \over r^\delta+r_\beta^\delta} \ ,
    \label{betaofrmodel}
\end{equation}
where $\beta_0$ and $\beta_\infty$ are the values of $\beta$ at $r=0$ and at infinity, respectively,  $r_\beta$ is the anisotropy radius where $\beta$ is midway beween $\beta_0$ and $\beta_\infty$, and $\delta$ is the anisotropy sharpness, with
$\delta=1$ for \citet{Tiret+07} anisotropy and $\delta=2$ for the generalized Osipkov-Merritt (`gOM') anisotropy (\citealt{Osipkov79,Merritt85-df}).
For $\delta = 1$ or 2, the exponential term in Eq.~(\ref{e:sigmar}) is (Appendix B of \citealt{MBB13} for these anisotropy models and a few others)
\[
\left({s^\delta+a^\delta \over s^\delta+a^\delta}\right)^{2\,(\beta_\infty-\beta_0)/\delta} \ .
\]

The anisotropy radius was either a  free parameter or fixed to the scale radius of the given morphology, $r_\nu$, previously fitted to the photometric data in \citetalias{Cava+17}, using a projected NFW model plus a constant field surface density.
In \citetalias{Mamon+19}, we found that the elliptical galaxy distribution traces the mass very well, the S0 distribution traces it reasonably well, while the spiral galaxy distribution traces it very poorly. In other words, $r_{\nu,{\rm E}} \simeq r_{-2}$, while $r_{\nu,{\rm S}}\simeq 4\,r_{-2}$.

\begin{table*}
\centering
\caption{Mass - velocity anisotropy models}
\label{t:models}
\setlength{\tabcolsep}{2pt}
\begin{tabular}{rc|ccccccc}
\hline
  \hline
   & Model & 6 & 7 & 7c & 12 & 12e & 15 & 15e \\
  \hline
  (1) & color & \textcolor{magenta}{magenta} & \textcolor{orange}{orange} & \textcolor{red}{red} & \textcolor{Turquoise}{turquoise} &  \textcolor{RoyalBlue}{royal blue} & \textcolor{ForestGreen}{green} & \textcolor{NavyBlue}{navy blue} \\
  (2) & $\rho(r)$ & gNFW & gNFW & gNFW & NFW & $n$=6 Einasto & NFW & $n$=6 Einasto \\
  (3) & $\beta(r)$ & gOM & Tiret & Tiret & Tiret & Tiret & gOM & gOM \\  
  (4) & $R^{-1}$ & 0.011 & 0.040 & 0.040 & 0.031 & 0.031 & 0.003 & 0.002 \\
  (5) & $N_{\rm free}$ & 12 & 15 & 15 & 14 & 14 & 11 & 11 \\
  (6) & $r_{200}$ & free & free & free & free & free & free & free \\
  (7) & $\cvir$ & $f(\mvir)$ & $f(\mvir)$ & free & $f(\mvir)$ & $f(\mvir)$ & $f(\mvir)$ & $f(\mvir)$ \\
  (8) & $\gamma$ & free & free & free & -- & -- & -- & -- \\
  \hline
  (9) & $\rnu$ & free & free & free & free & free & free & free \\
  (10) & $\beta_0$ & free & free & free & free & free & free & free \\
  (11) & $\beta_\infty$ & free & free & free & free & free & free & free \\
  (12) & $\rb $ & $\rnu$ & free & free & free & free & $\rnu$ & $\rnu$ \\
  \hline
  \end{tabular}
  \tablefoot{The model number is the same as in table~2 of \citetalias{Mamon+19}. Letters following the model numbers indicate slight modifications to the models; we use `c' to indicate that the halo concentration $\cvir$ is a fully free parameter, and `e' that the Einasto total density model $\rho(r)$ model is adopted. 
  The rows are 
(1): color coding used in the figures (unless otherwise specified); 
  (2): total density model;
  (3): velocity anisotropy model (parameter $\delta$ of Eq.~[\ref{betaofrmodel}]);
  (4): \cite{Gelman&Rubin92} convergence criterion of the 6 MCMC chains (values below 0.02 are considered very good, and values between 0.02 and 0.04 are considered adequate);
  (5): number of free parameters;
  (6): virial radius;
  (7): halo concentration (when $\cvir=f(\mvir)$ we use Eq.~[\ref{cofMMaccio}] from \citet{DM14} to estimate $\cvir$ from $\mvir$, so $\cvir$ is not a fully free parameter: it has a Gaussian prior with uncertainty $\sigma(\log c) = 0.1$);
  (8): slope of the inner total mass density profile;
  (9--12): for each of the three morphological classes, the scale radius of the number density profile (9), the inner (10) and outer (11) velocity anisotropies, and the anisotropy radius (12), where $\rb=\rnu$ means that we fixed the anisotropy radius to the tracer scale radius.
}
\end{table*}

In the present paper, among the 30 models of \citetalias{Mamon+19}, we only considered single-component mass models with free inner and outer anisotropy for all three morphological types. We also excluded the models with Tiret anisotropy with anisotropy radius fixed to $r_\beta=r_\nu$, which lead to linear  $\beta-\gamma$ relations  if tracer follows mass (Mamon \& Biviano, in prep.). This left us with models 6, 7, 12, and 15 in Table~2 of \citetalias{Mamon+19}. Our results are therefore independent from the linear  $\beta-\gamma$ relation assumption that according to some authors could explain the power-law behavior of $Q(r)$ and $\qr(r)$ \citep{DML05,AW20}. 
  \begin{figure}
   \centering
   \includegraphics[width=\hsize]{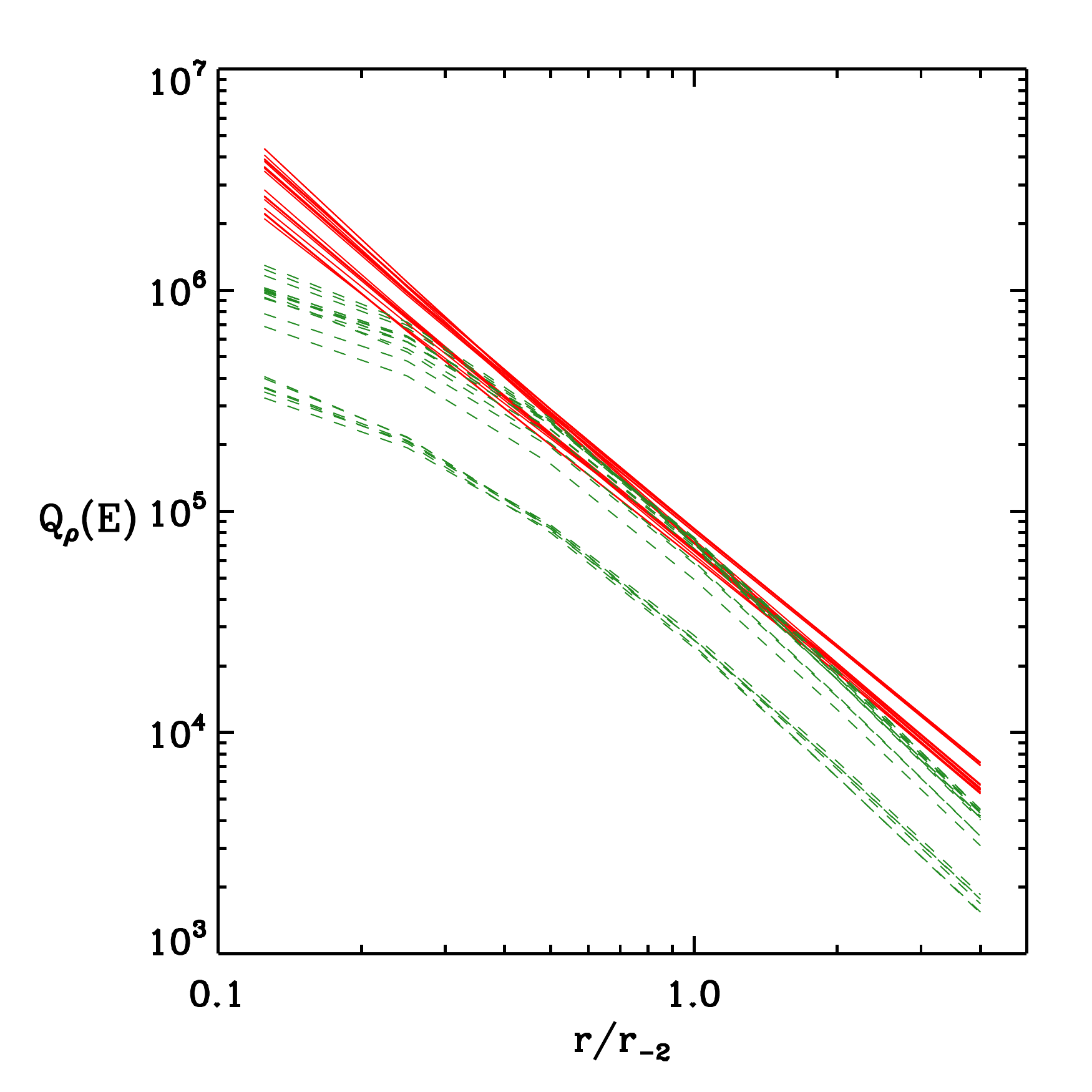}
      \caption{Examples of 20 linear, as defined by $l = 1-D/L \geq 0.9$ (\emph{red solid lines}), and 20 non-linear, $l<0.9$ (\emph{green dashed lines}), $Q_\rho(r)$ for ellipticals in model~7. $L$ and $D$ are defined in Eqs.~(\ref{e:Len}) and (\ref{e:Dev}), respectively.
      } 
         \label{f:lin}
   \end{figure}

All four considered \citetalias{Mamon+19} models assume NFW $\nu(r)$, with a scale radius $\rnu$ as a free parameter, one $\rnu$ parameter for each morphological class. In models 6 and 7, $\rho(r)$ is modelled by the 
gNFW profile, with $\gamma_{\infty}=-3$ and $\gamma_0$ as a free parameter (Eq.~[\ref{rhoofrmodel}]). In models 12 and 15, $\rho(r)$ is instead modelled by the NFW profile. In all four models $\rvir$, and therefore $\mvir$, is a free parameter of $\rho(r)$, while $\cvir$ is related to $\mvir$ through the relation of 
\citet{DM14}: 
\begin{equation}
    \log \cvir= 2.13-0.10 \,\log \left({\mvir\over \ms}\right) \ ,
    \label{cofMMaccio}
\end{equation}
with a Gaussian prior $\sigma(\log c_{200}) = 0.1$.\footnote{The logarithms are all in base 10.}
Therefore the mass density profile involves 2 (NFW and $n$=6 Einasto) or 3 (gNFW) free parameters.

Models~7 and 12 adopt the 
Tiret model for $\beta(r)$,
while models~6 and 15 adopt the
gOM anisotropy model.
Both the Tiret and the gOM models are characterized by two free parameters per each morphological class, the inner and outer velocity anisotropies $\beta_0$ and $\beta_{\infty}$. The anisotropy scale radius $\rb$ is a free parameter in Tiret models 7 and 12, whereas it is tied to the tracer scale radius, $\rb=\rnu$ in gOM models 6 and 15. 
Thus, the anisotropy profile involves 2 (fixed $r_\beta$) to 3 (free $r_\beta$) parameters per morphological type, hence 6 (fixed $r_\beta$) to 9 (free $r_\beta$) free parameters after summing over the three morphological types.

In addition to the four models described above we consider the three following models. Model 7c is the same as model 7 but with $\cvir$ as a fully free parameter (with a uniform prior for  $\log c$ from 0 to 1). Model 12e and  15e are the same as, respectively, model 12 and 15, but with a $n=6$ Einasto $\rho(r)$ in lieu of NFW. The properties of these seven models are summarized in Table~\ref{t:models}.

For each MCMC chain element, we determine $Q(r)$ and $Q_r(r)$ at six logarithmically spaced radii, from $r/\rtwo = 0.125$ to 4, in steps of a factor 2: \mbox{$r/\rtwo=2^{i-4}$, $i=1, \ldots, 6$}, that is from roughly 0.03 to 1 virial radius. We fit straight lines to the six values of $\log Q$ vs. $\log r$, $\log Q_r$ vs. $\log r$, for each individual MCMC, yielding $\log Q(r) = a + b\,\log (r/r_{-2})$. We measure the linearity of $Q(r)$ and $Q_r(r)$  using the quantity $l=1-D/L$, 
where $L$ is the length of the fitted line,
\begin{equation}
L \equiv (1+b^2)^{1/2} \, \mid x_6 - x_1 \mid \ ,
\label{e:Len}
\end{equation}
where $x_i = \log(r_i/r_{-2})$,
and $D$
is the orthogonal deviation of the six measurements from the fitted line,
\begin{equation}
D \equiv  (1+b^2)^{-1/2}\,\sum_{i=1}^6 \, \mid y_i - (b \ x_i + a) \mid  \ ,
\label{e:Dev}
\end{equation}
where $y_i = \log[Q(r)/Q(r_{-2})]$, or its analog for $Q_r$ instead of $Q$.
We arbitrarily set a limit $l=0.9$ above which the relation is considered to be linear, that is the points deviate on average from the fitted line by less than 10\% of the line length. We show examples of linear and non-linear relations in Fig.~\ref{f:lin}.

\section{Results}
\label{s:res}
\subsection{Linearity of $\log Q$ vs. $\log r$}
\label{s:linearity}
\begin{figure}
   \centering
   \includegraphics[width=\hsize]{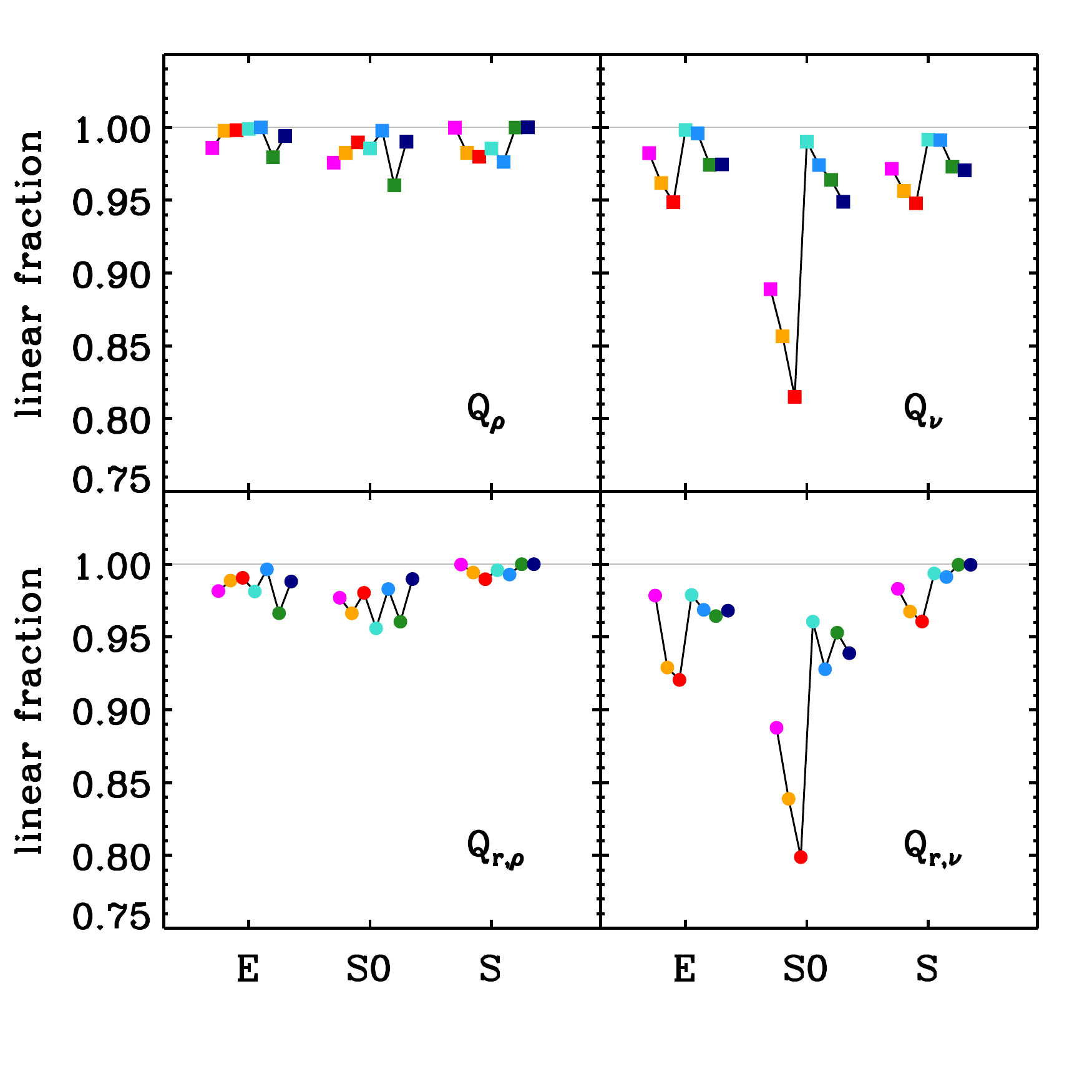}
      \caption{Fraction of linear, $l \geq 0.9$, relations deduced from  the MCMC chain elements, for E, S0, and S tracers, in different models (color coded as in Table~\ref{t:models}) for the {\tt sigv} scaling, using total mass density profile (\emph{left}) and tracer number density profiles (\emph{right}), with total velocity dispersion (\emph{top}) and radial velocity dispersion (\emph{bottom}). Error bars are smaller than the symbol sizes.}
         \label{f:linmod}
   \end{figure}

We first consider whether the $Q(r)$ and $Q_r(r)$ profiles are linear in logarithmic space.  Fig.~\ref{f:linmod} shows the fraction $f_l$ of MCMC chain elements that have $l \geq 0.9$ (see Sect.~\ref{s:ana}) with the $f_l$ values listed in Table~\ref{t:qqr}. Independently of the chosen model and galaxy type, all profiles are linear for over 95\% of the MCMC chains for $\qrho$ and $\qrrho$. The $f_l$ values of the $\qrho$ and $\qrrho$ profiles are almost identical. There is no clear dependence of $f_l$ on either the $\rho(r)$  or the $\beta(r)$ model chosen. Recall that we did not consider the models of \citetalias{Mamon+19} that lead to linear $\beta-\gamma$ relations to avoid biasing the linearity of the PPSD, since the PPSD and $\beta-\gamma$ relations may be physically related.  
 
The $\qnu$ and $\qrnu$ profiles are also linear for over 90\% of the MCMC chain elements, independently of the chosen model, but only when either ellipticals or  spirals are considered. When considering S0, the $f_l$ values for the $\qnu$ and $\qrnu$ profiles can be as low as \mbox{$\simeq  0.80$}. Models with gNFW  $\rho(r)$ have lower values of $f_l$ when considering S0. The $f_l$ values of the $\qnu$ and $\qrnu$ profiles are very similar. 

Combining all three morphological classes, the linear fractions for $\qrho$ and $\qrrho$ are maximal for model 15e ($n=6$ Einasto with gOM anisotropy).
Similarly, for $\qnu$ and $\qrnu$, the linear fractions are maximal for model 12 (NFW with Tiret anisotropy).

\begin{table*}
\centering
\caption{$Q$ and $Q_r$ profiles: $f_l$ and slopes for the \texttt{sigv} scaling}
\label{t:qqr}
\resizebox{\textwidth}{!}{
\begin{tabular}{r|cc|cc|cc||cc|cc|cc}
  \hline \\
   & \multicolumn{6}{c}{$\qrho$} & \multicolumn{6}{c}{$\qnu$} \\ \\
        & \multicolumn{2}{c}{E} & \multicolumn{2}{c}{S0} & \multicolumn{2}{c}{S} & \multicolumn{2}{c}{E} & \multicolumn{2}{c}{S0} & \multicolumn{2}{c}{S} \\
        Model &  $f_l$ & slope & $f_l$ & slope & $f_l$ & slope &  $f_l$ & slope & $f_l$ & slope & $f_l$ & slope \\
        \hline
6    &  0.99  & $-1.90 \pm  0.09 $ &  0.98  & $-1.76 \pm  0.10 $ &  1.00  & $-1.78 \pm  0.11 $ &  0.98  & $-1.76 \pm  0.10 $ &  0.89  & $-1.60 \pm  0.10 $ &  0.97  & $-1.24 \pm  0.11$ \\
 7    &  1.00  & $-1.85 \pm  0.09 $ &  0.98  & $-1.73 \pm  0.10 $ &  0.98  & $-1.83 \pm  0.18 $ &  0.96  & $-1.78 \pm  0.12 $ &  0.86  & $-1.63 \pm  0.10 $ &  0.96  & $-1.24 \pm  0.11$ \\
 7c   &  1.00  & $-1.81 \pm  0.08 $ &  0.99  & $-1.70 \pm  0.08 $ &  0.98  & $-1.73 \pm  0.17 $ &  0.95  & $-1.78 \pm  0.11 $ &  0.81  & $-1.64 \pm  0.10 $ &  0.95  & $-1.28 \pm  0.11$ \\
12    &  1.00  & $-1.84 \pm  0.09 $ &  0.99  & $-1.72 \pm  0.10 $ &  0.99  & $-1.82 \pm  0.18 $ &  1.00  & $-1.85 \pm  0.11 $ &  0.99  & $-1.67 \pm  0.10 $ &  0.99  & $-1.21 \pm  0.11$ \\
12e   &  1.00  & $-1.86 \pm  0.08 $ &  1.00  & $-1.74 \pm  0.09 $ &  0.98  & $-1.83 \pm  0.18 $ &  1.00  & $-1.81 \pm  0.10 $ &  0.97  & $-1.64 \pm  0.09 $ &  0.99  & $-1.22 \pm  0.11$ \\
15    &  0.98  & $-1.86 \pm  0.11 $ &  0.96  & $-1.73 \pm  0.11 $ &  1.00  & $-1.82 \pm  0.13 $ &  0.97  & $-1.84 \pm  0.12 $ &  0.96  & $-1.66 \pm  0.12 $ &  0.97  & $-1.21 \pm  0.11$ \\
15e   &  0.99  & $-1.89 \pm  0.10 $ &  0.99  & $-1.75 \pm  0.11 $ &  1.00  & $-1.82 \pm  0.12 $ &  0.97  & $-1.82 \pm  0.11 $ &  0.95  & $-1.65 \pm  0.11 $ &  0.97  & $-1.22 \pm  0.11$ \\
& & & & & & & & & & & \\
 mean & 0.99 & $-1.86 \pm 0.03$ & 0.98 & $-1.73 \pm 0.02$ & 0.99 & $-1.80 \pm 0.03$ & 0.97 & $-1.81 \pm 0.03$ & 0.92 & $-1.64 \pm 0.02$ & 0.97 & $-1.23 \pm 0.02$ \\
\hline \\
   & \multicolumn{6}{c}{$\qrrho$} & \multicolumn{6}{c}{$\qrnu$} \\ \\
        & \multicolumn{2}{c}{E} & \multicolumn{2}{c}{S0} & \multicolumn{2}{c}{S} & \multicolumn{2}{c}{E} & \multicolumn{2}{c}{S0} & \multicolumn{2}{c}{S} \\
  Model      &  $f_l$ & slope & $f_l$ & slope & $f_l$ & slope &  $f_l$ & slope & $f_l$ & slope & $f_l$ & slope \\
\hline
 6    &  0.98  & $-2.09 \pm  0.23 $ &  0.98  & $-1.85 \pm  0.20 $ &  1.00  & $-1.88 \pm  0.14 $ &  0.98  & $-1.98 \pm  0.20 $ &  0.89  & $-1.73 \pm  0.17 $ &  0.98  & $-1.48 \pm  0.14$ \\
 7    &  0.99  & $-1.92 \pm  0.21 $ &  0.97  & $-1.77 \pm  0.20 $ &  0.99  & $-2.03 \pm  0.32 $ &  0.93  & $-1.88 \pm  0.20 $ &  0.84  & $-1.70 \pm  0.18 $ &  0.97  & $-1.46 \pm  0.17$ \\
 7c   &  0.99  & $-1.90 \pm  0.19 $ &  0.98  & $-1.75 \pm  0.16 $ &  0.99  & $-1.91 \pm  0.29 $ &  0.92  & $-1.89 \pm  0.20 $ &  0.80  & $-1.70 \pm  0.17 $ &  0.96  & $-1.46 \pm  0.17$ \\
12    &  0.98  & $-1.86 \pm  0.17 $ &  0.96  & $-1.73 \pm  0.17 $ &  1.00  & $-1.99 \pm  0.28 $ &  0.98  & $-1.87 \pm  0.18 $ &  0.96  & $-1.68 \pm  0.17 $ &  0.99  & $-1.41 \pm  0.16$ \\
12e   &  1.00  & $-1.89 \pm  0.17 $ &  0.98  & $-1.75 \pm  0.18 $ &  0.99  & $-2.02 \pm  0.31 $ &  0.97  & $-1.85 \pm  0.18 $ &  0.93  & $-1.67 \pm  0.16 $ &  0.99  & $-1.43 \pm  0.18$ \\
15    &  0.97  & $-1.90 \pm  0.21 $ &  0.96  & $-1.75 \pm  0.19 $ &  1.00  & $-1.92 \pm  0.16 $ &  0.96  & $-1.89 \pm  0.22 $ &  0.95  & $-1.68 \pm  0.19 $ &  1.00  & $-1.44 \pm  0.15$ \\
15e   &  0.99  & $-1.98 \pm  0.21 $ &  0.99  & $-1.80 \pm  0.19 $ &  1.00  & $-1.93 \pm  0.15 $ &  0.97  & $-1.92 \pm  0.21 $ &  0.94  & $-1.71 \pm  0.19 $ &  1.00  & $-1.45 \pm  0.15$ \\
& & & & & & & & & & & \\
mean & 0.98 & $-1.93 \pm 0.08$ & 0.97 & $-1.77 \pm 0.04$ & 1.00 & $-1.94 \pm 0.06$ & 0.96 & $-1.90 \pm 0.04$ & 0.90 & $-1.70 \pm 0.02$ & 0.99 & $-1.45 \pm 0.02$ \\
\hline
\end{tabular}}
\tablefoot{Columns labelled '$f_l$' give the fraction of linear ($l=1-D/L\geq 0.9$) MCMC $Q$ profiles. Columns labelled `slope' give the average and dispersion of the slopes of the MCMC $Q$ profiles with $l \geq 0.9$. Rows labelled `mean' list the weighted mean and dispersion of all the model slopes, using the inverse of model slope dispersions as weights.}
\end{table*}

    \begin{figure}
   \centering
   \includegraphics[width=\hsize]{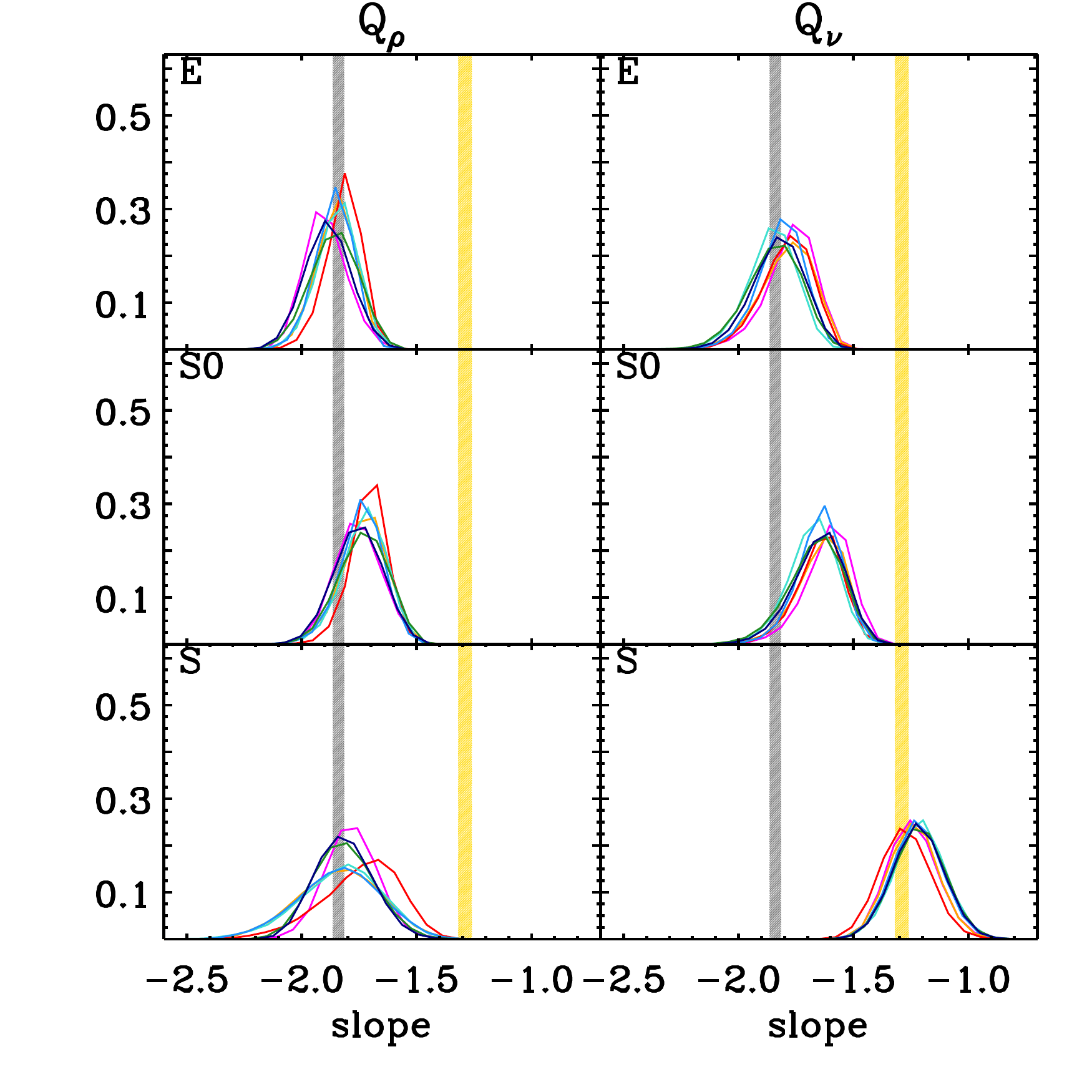}
      \caption{Marginal distributions of the logarithmic slopes of the linear ($l \geq 0.9$) $Q$ profiles, for different morphological classes (E, S0, S in the \emph{top}, \emph{middle}, and \emph{bottom} panel, respectively) using different models (color coded as in Table~\ref{t:models}), for \texttt{sigv} scaling. Left panels: $\qrho$; right panels: $\qnu$. \emph{Grey} (respectively \emph{yellow)} \emph{shadings} indicate the simulation-based prediction for the slope of DM tracers (respectively subhalos), $-1.84 \pm 0.025$ (respectively $-1.29 \pm 0.03$). 
     }
         \label{f:qdist}
   \end{figure}
   
      \begin{figure}
   \centering
   \includegraphics[width=\hsize]{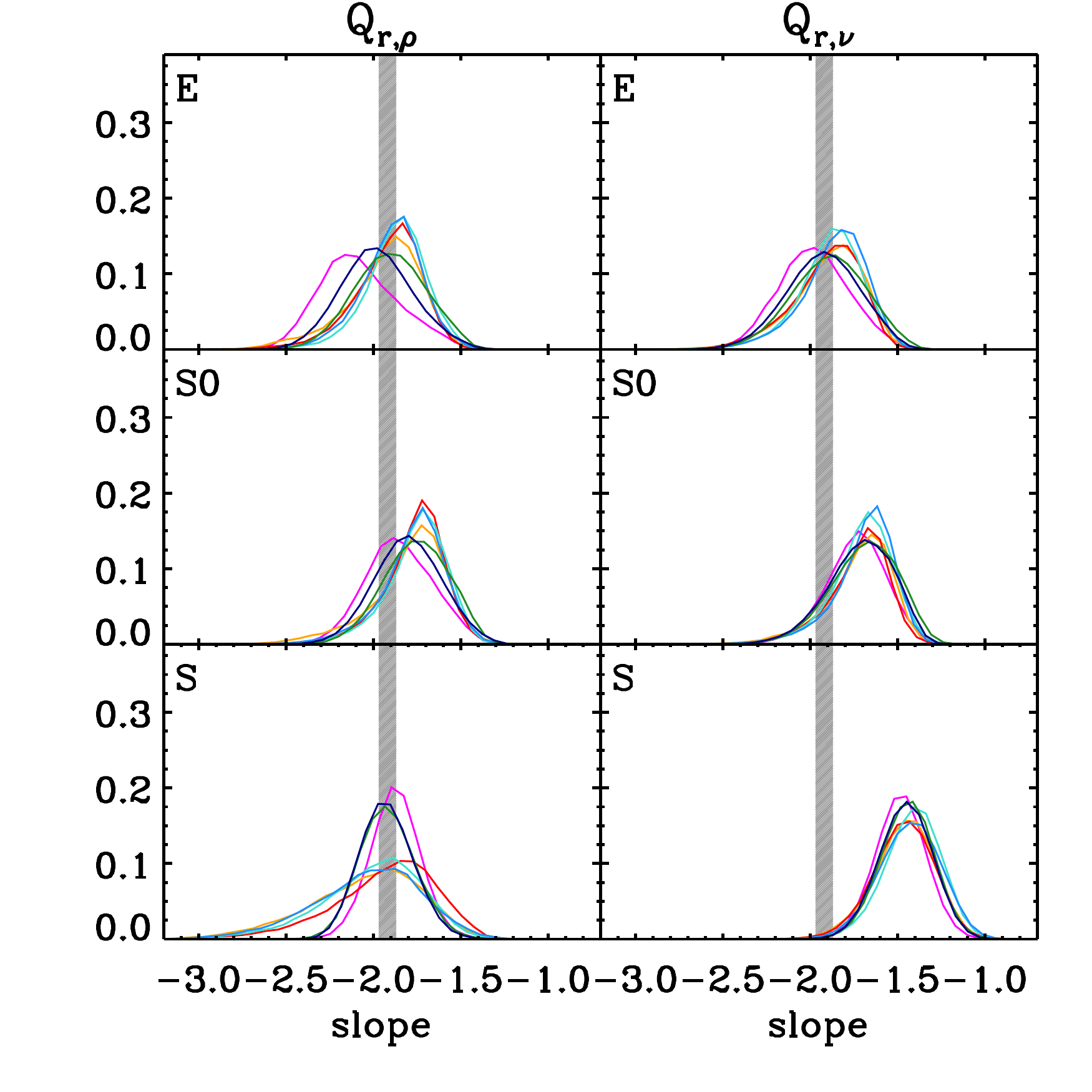}
      \caption{Same as Fig.~\ref{f:qdist} but for $Q_r$ instead of $Q$. \emph{Grey shadings} indicate the simulation-based prediction for the slope of DM tracers, $-1.92 \pm 0.05$.}
         \label{f:qrdist}
   \end{figure}

 \subsection{Slopes}
 \label{s:slopes}
 We then fit straight lines to $\log Q$ and $\log Q_r$ vs. $\log(r/\rtwo)$, for  the MCMC chain elements for which $l \geq 0.9$ (non linear profiles are not considered as the straight line slope is not a useful statistic for them).
 We show the distributions of the best-fit logarithmic slopes of $Q(r)$ in Fig.~\ref{f:qdist} (left panel: $\qrho$, right panel: $\qnu$) and of $Q_r(r)$ in Fig.~\ref{f:qrdist}.  The slope distributions do not differ in a significant way from one model to another and have similar unimodal shapes for all profiles. 
 
We compare our observational results with the predictions for DM particles from cosmological simulations, adopting the slope values that \citet{DML05} obtained from the DM-only cosmological simulations of \citet{Diemand+04a,Diemand+04b}, $-1.84$ and $-1.92$ for $Q(r)$ and $\qr(r)$, respectively, with uncertainties of 0.025 and 0.05, respectively, to account for the scatter among the values found in different studies, that include both DM-only and hydrodynamical simulations \citep{TN01,RTM04,DML05,KKH08}. We also compare the observational results with the only available predictions for subhalos in cosmological hydrodynamical simulations, those of \citet{Marini+21}, who found a $Q(r)$ slope of $-1.29 \pm 0.03$ (the authors did not study $Q_r(r)$).

\begin{figure}
   \centering
   \includegraphics[width=\hsize]{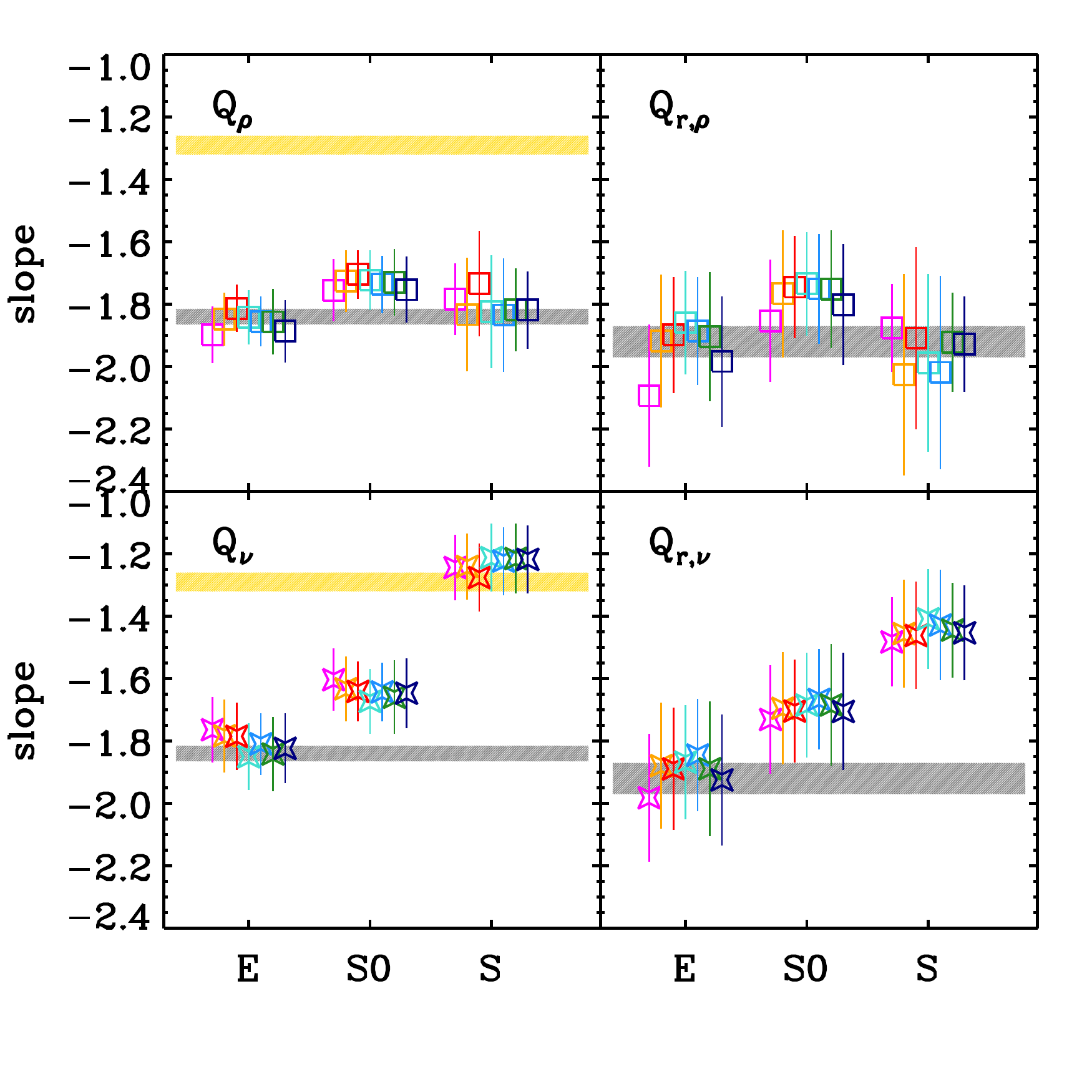}
      \caption{Average and dispersion of  the $Q$ profile logarithmic slope and the simulation-based predictions for DM tracers (\emph{grey shading}), $-1.84 \pm 0.025$ for $Q(r)$ and $-1.92 \pm 0.05$ for $Q_r(r)$, and for subhalos (\emph{yellow shading}), $-1.29 \pm 0.03$, for different morphological classes (indicated on the $x$ axis), in different models (color coded as in Fig.~\ref{f:linmod} and Table~\ref{t:models}). Only linear ($l \geq 0.9$) profiles are considered.} 
         \label{f:qdev}
   \end{figure}

 Fig.~\ref{f:qdev}  shows the biweight means and dispersions of the marginal distributions of the PPSD slopes (total and radial) shown in Figs.~\ref{f:qdist} and \ref{f:qrdist}, compared with the simulation-based values. 
 We also provide the average and dispersion of the logarithmic slopes of the linear $Q, Q_r$ profiles for all models and all galaxy types in Table~\ref{t:qqr}. Our results do not depend in a significant way on the assumed model for $\rho(r)$ and $\beta(r)$. In fact, the average logarithmic slopes of the $Q$ and $Q_r$ profiles for a given galaxy type are very similar across different models, and the dispersions of the average slope values of the seven models is much smaller than the dispersion in the values of the  slopes obtained from the MCMC of any individual model  (see rows labelled `mean' in Table~\ref{t:qqr}). 
 
Both for ellipticals and spirals, and also marginally for S0s, the logarithmic slopes of the linear $\qrho$ and $\qrrho$  profiles  are consistent with the simulation-based prediction for DM tracers for all models  (Fig.~\ref{f:qdev}). The $\qnu$ and $\qrnu$ profile slopes for ellipticals are consistent with those of simulations based on DM tracers,
while the corresponding slopes for spirals are not.
The $\qnu$ slopes for S0s are also inconsistent with the simulation-based predictions based on DM tracers for all models, while the $\qrnu$ profile slopes for S0s are marginally consistent with the same simulation predictions (thanks to larger dispersions).  Interestingly, the spiral
$\qnu$ profile slopes are in agreement with the simulation-based prediction for subhalos, while the elliptical and S0 $\qnu$ profiles are not.

\section{Discussion} \label{s:dis}

We discuss in turn our results on {\tt sigv} stacks and on the other two stacks ({\tt num} and {tempX}).

\subsection{Discussion of results on {\tt sigv} stacked clusters}
\label{ss:sigv}
Most $\qrho$ and $\qrrho$ profiles are very close to power-law relations: 96\% of all models and galaxy types show PPSDs with linearity $l>0.9$
(see Fig.~\ref{f:linmod}, left panels, and Table~\ref{t:qqr}). The large majority of MCMC chain elements predict power-law $\qrho(r)$ and $\qrrho(r)$ with average slopes in very good agreement and fully consistent with the simulation-based expectations using DM particles as tracers,
but slightly flatter for S0s than for ellipticals and spirals (see Fig.~\ref{f:qdev}, top-left panel). Our results support the findings of several studies based on both DM-only and hydrodynamical simulations \citep{TN01,RTM04,DML05,Ludlow+10,Navarro+10}, and of previous observational studies \citep{Biviano+13,MBM14,Biviano+16,Biviano+21}, and do not support claims against the power-law behavior of $Q(r)$ \citep{NOJ17}. Since our results are based on a stack cluster, we can neither confirm nor reject the numerical result of \citet{Schmidt+08} against the universality of $Q(r)$ across different cosmological halos. However, for none of the three galaxy classes do the $\qrho(r)$ slopes agree with those obtained for subhalos in numerical simulations \citep{Marini+21}.

If both $Q(r)$ and $Q_r(r)$ are power laws, of respective slopes $\alpha$ and $\alpha+\Delta \alpha$, then their ratio ${\cal R} = Q_r/Q = (3-2\,\beta)^{3/2}$ should also be a power law of slope $\Delta \alpha$.
For the Tiret and gOM anisotropy models (Eq.~[\ref{betaofrmodel}]), one then expects
\begin{equation}
{\cal R}(y) =  3-2\,\beta_0 - 2\,\left(\beta_\infty-\beta_0\right)\,
{y^\delta\over y^\delta+1} \propto y^{2\Delta\alpha/3} \ ,
\label{fofy1}
\end{equation}
where 
$y=r/r_\beta$.
Eq.~(\ref{fofy1}) indicates that  ${\cal R}$ varies from one constant value, ${\cal R}_0 = 3-2\,\beta_0$, at small radii, to another constant value, ${\cal R}_\infty = 3-2\,\beta_\infty$, at large radii.
Therefore, $Q_r/Q$ cannot be a power law over the full range of radii (unless $\beta_\infty=\beta_0$).
\begin{figure}
    \centering
  \includegraphics[width=\hsize]{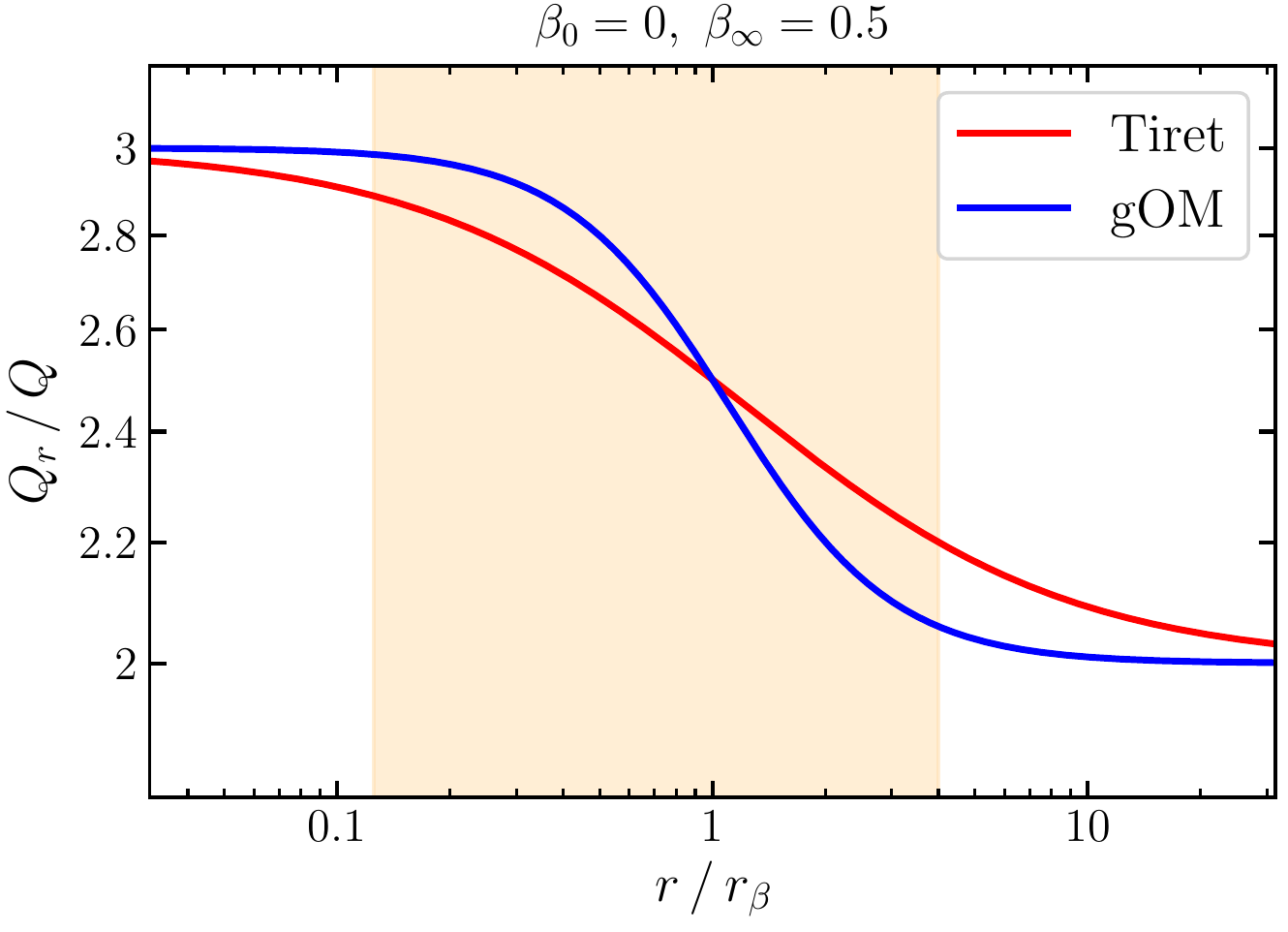}
  \caption{Illustration of the non-linearity of $Q_r/Q$ in Tiret and gOM anisotropy models (Eq.~[\ref{fofy1}] with $\delta=1$ and 2, respectively). Our analysis was limited to the radii in the shaded region.}
    \label{fig:fofy}
\end{figure}
If one restricts the analysis to a narrow range of radii around $r=r_\beta$, one expects a quasi-linear behavior obtained by a series expansion of ${\cal R}(y)$ in Eq.~(\ref{fofy1}): 
\begin{equation}
    {\cal R}(y) = 3 - \beta_0 - \beta_\infty
-{1\over 2}\,\left(\beta_\infty-\beta_0\right)\,\delta\,(y-1)
+ o(y-1) \ .
\label{fofyseries}
\end{equation}
The zeroth order term is positive since $\beta<1$ by definition.
The first order term is proportional to $\delta$, and is negative for $\beta_\infty>\beta_0$ but positive otherwise.
Hence, the transition of $Q_r/Q$ from ${\cal R}_0$ at small radii to ${\cal R}_\infty$ at large radii is smoother for low $\delta$ anisotropy profiles. This is illustrated in Figure~\ref{fig:fofy}, which shows that the
gOM model ($\delta=2$) is less linear than  the Tiret ($\delta=1$) model.
 In turn, this would indicate that the fraction of linear models should be higher with Tiret anisotropy than for similar mass models with gOM anisotropy. However, in practice, the necessary non-linearity of $Q_r/Q$ is not a worry, because the non-linearity range of $Q_r/Q$ is smaller than the non-linearity range of either $Q(r)$ or $Q_r(r)$, because the logarithmic slopes of $Q(r)$ and $Q_r(r)$ are similar (Table~\ref{t:qqr}). For example, if $Q(r)$ were perfectly linear ($l=1$), then $Q_r$ would have a linearity $l = 0.997$ and 0.992 for the Tiret and gOM anisotropy models, respectively, hence much greater than our threshold of 0.9 for linear models. 

It might at first appear surprising that $\qrho$ and $\qrrho$ should have similar slopes for the three morphological classes, given that the three classes have different line-of-sight velocity dispersion profiles \citepalias{Cava+17} and different $\beta(r)$ \citepalias{Mamon+19}.
The similarity of $\qrho$ and $\qrrho$ for the three classes then imply that they also have similar $\sigma(r)$ and $\sigma_r(r)$ and that the observed differences in their line-of-sight velocity dispersion profiles \citepalias{Cava+17} and $\beta(r)$ \citepalias{Mamon+19} is compensated by their different $\nu(r)$ (see Eqs.~[\ref{e:sigmar}], [\ref{e:sig}], and \citetalias{Cava+17}).

One expects larger differences between the $Q_\nu(r)$ profiles of different morphological classes, because $Q_\nu$ is proportional to the number density of that class, and the number concentrations of the best-fit NFW models of each class differ significantly \citepalias{Cava+17}.
Indeed, the PPSDs of $\qnu$ and $\qrnu$ are increasingly shallower when moving from ellipticals to S0s to spirals (bottom panels of Fig.~\ref{f:qdev}), even if these profiles are also quite close to power-law relations, with $f_l \gtrsim 0.8$ for all models and all galaxy types (see Fig.~\ref{f:linmod}, right panels, and Table~\ref{t:qqr}). At variance with $\qrho$ and $\qrrho$, only for ellipticals is there a good agreement between the observed $\qnu$ and $\qrnu$ slopes and the expected values from simulations using DM particles as tracers (bottom panels of Fig.~\ref{f:qdev}). This is not surprising, given that $\nu(r) \approx \rho(r)$ for ellipticals, but not for the other two types \citepalias{Mamon+19}. 

Interestingly, the logarithmic slope of $\qnu(r)$  for spirals is very similar to the one found by \citet{Marini+21} for subhalos in cluster-size halos in cosmological hydrodynamical simulations (see bottom-left panel of Fig.~\ref{f:qdev}). 
This similarity is probably related to the more extended radial distributions of spirals on one hand \citepalias{Cava+17} and of subhalos on the other (\citeauthor{Marini+21}). 
Note that subhalos in dark matter only cosmological simulations of the same resolution show instead that the power-law dynamical entropy turns to flat inside half a virial radius.

The more extended subhalo number density profile, if not due to numerical effects \citep{vandenBosch&Ogiya18}, can be explained in several ways.
Strong cluster tides at pericenter 
remove mass from infalling subhalos \citep{Merritt83},  as seen in simulations
(e.g. \citealt{Hayashi+03,Saro+06,Springel+08}). Note again that the steeper dynamical entropy (hence steeper $Q_\nu$) for the subhalos in hydrodynamical simulations relative to those in dark matter only ones suggests that the dissipative nature of gas leads to more concentrated subhalos that are more resilient to cluster tides. 
Such tides will remove mass from those 
subhalos that traverse the inner regions of clusters, causing (some of) the galaxies associated to them to fall below the data luminosity threshold. But tides affect all classes of galaxies, not just spirals. 
Alternatively, ram pressure stripping of the gas of spiral galaxies will strangle their subsequent star formation, leading to lower luminosities than gas-poor galaxies with the same orbits \citep{GG72,Boselli+16}. Another explanation may lie in temporal segregation instead of spatial segregation. If spiral galaxies are rapidly transformed into S0s and progressively into ellipticals (as argued, e.g. in \citetalias{Mamon+19}), then S0s and ellipticals are the end products of galaxies that entered earlier in the cluster, most probably from lower apocenters. Thus the radial distribution of spirals is much more extended than S0s and ellipticals, leading to the shallower $Q_\nu$ slope of spirals.
However, spirals are not expected to be the  dominant morphological class in simulated cluster subhalos. Therefore, the good agreement between the $Q_\nu$ slopes of observed spirals and simulated subhalos remains an open question.

Our results for the $\qnu$ and $\qrnu$ profiles agree with those obtained from analysis of observations of \citet{Capasso+19}, who determined $\qnu(r)$ for passive galaxies only, but not with \citet{MBM15} and \citet{AADDV17}, who found $\qnu(r)$ to be consistent with the simulation-based expectations by \citet{DML05}, for {\it all} classes of galaxies in two nearby clusters. Perhaps, thanks to our large data set, we are able to detect significant differences that were not visible in individual cluster analyses because of limited statistics.

When comparing  $\qrho, \qrrho$ versus $\qnu, \qrnu$, we should take into account that we forced the NFW model for $\nu(r)$, but allowed three different models for $\rho(r)$ (see Table~\ref{t:models}). However, our results are very insensitive to the choice of the $\rho(r)$ model, and models 12 and 15, that use NFW for $\rho(r)$, behave very similarly to all the others. Our analysis then suggests that the $\rho$-based definition of $Q$ and $\qr$ is more fundamental than that based on $\nu$, even if, observationally, $\qrho$ and $\qrrho$ are derived using inhomogeneous quantities, as $\rho(r)$ refers to the distribution of total matter, dominated by DM, and $\sigma, \sigma_r$ to the velocity dispersion of galaxies.

To interpret our results, we note that recent numerical simulations \citep{Colombi21} show that the power-law $\qrho$ and $\qrrho$ profiles are established very early on for cluster DM, during the violent relaxation phase, possibly because of a tendency of the system towards a state of minimal entropy \citep{HK10}. As galaxies enter the cluster gravitational potential well, their orbits and spatial distributions may evolve to reach the same state of dynamical entropy ($\propto Q^{-2.3}$), leading to the same $\qrho$ and $\qrrho$ as that of DM. Since the spatial distribution of ellipticals is similar to that of the total matter, we argue that the bulk motions of ellipticals  experienced the same process of violent relaxation as the total matter, that is their progenitors (perhaps with different morphologies)  were present at the time of cluster formation. 

Violent relaxation at cluster formation cannot be the process shaping the PPSD profiles of S0s and spirals. Spirals have probably entered the cluster within the last $\sim 2$ to 3 Gyr, after which they are morphologically transformed to S0s and/or ellipticals (e.g. \citealt{LTC80}; \citealt{Couch+98}; see also \citetalias{Mamon+19}), and quenched by the cluster environment \citep[e.g.][]{Poggianti+04,Haines+13}, with indications that morphological transformation precedes star formation quenching \citep{Sampaio+22}. There is also observational evidence that S0s are not a pristine cluster population \citep{Postman+05,Smith+05,Desai+07}. The deviation of the  S0s and spirals $\qnu$ and $\qrnu$ profiles from simulation-based expectations for DM particles is probably an indication that their PPSD is achieved in a different way from ellipticals. S0s are an intermediate population between that of ellipticals and spirals, in terms of their PPSD. If S0s originate from spirals through some environmental process, such a process could also be responsible for the gradual PPSD evolution from that of spirals to that of ellipticals \citepalias{Cava+17}. However, no such evolution is seen for the subhalo PPSD in cluster-sized halos from cosmological simulation
(I. Marini, priv. comm.).

  \begin{figure}
   \centering
   \includegraphics[width=\hsize]{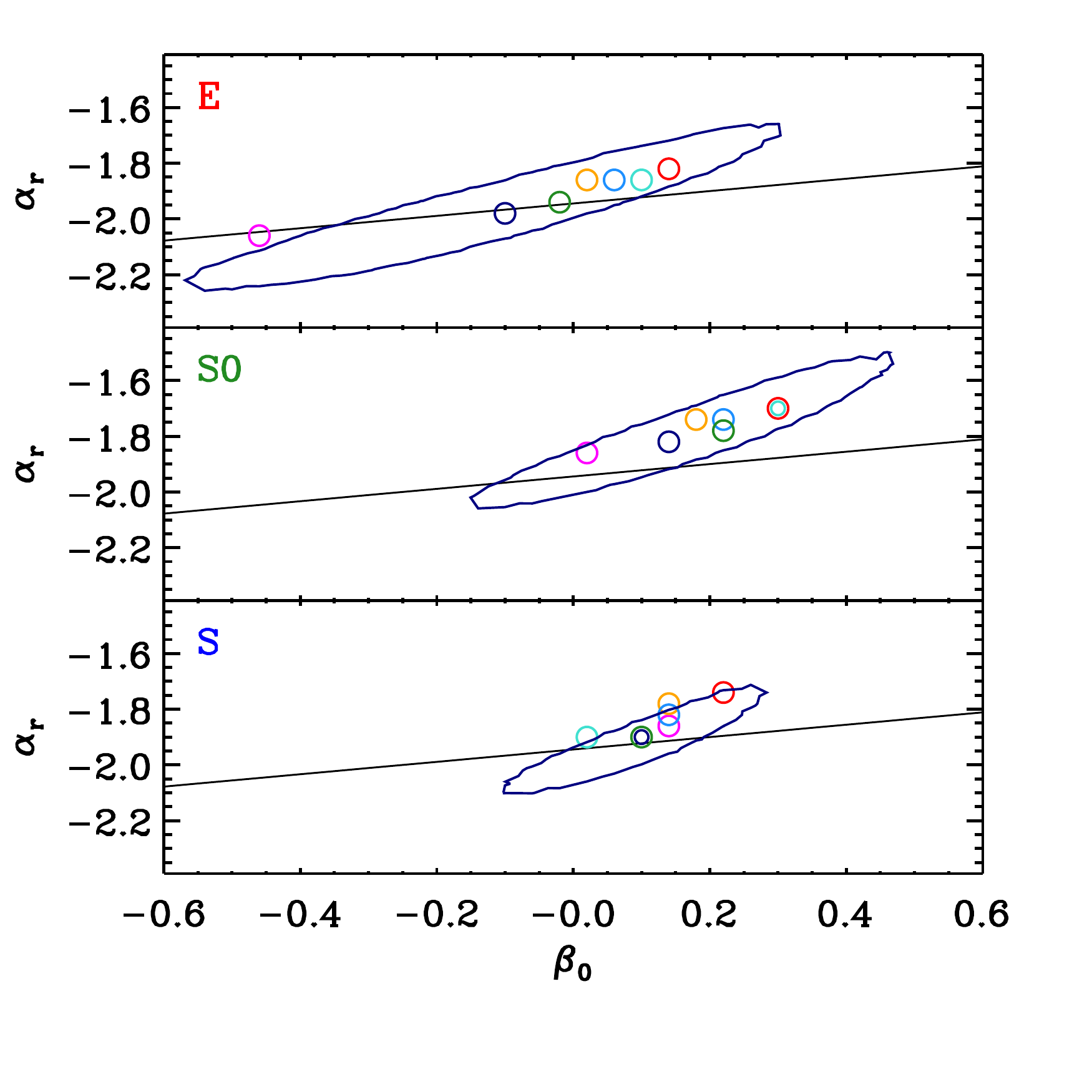}
   \caption{Distributions of $Q_{r,\rho}$ logarithmic slopes vs. central velocity anisotropy, $\beta_0$ for the E (\emph{top panel}), S0 (\emph{middle panel}), and S (\emph{bottom panel}) classes. 
   \emph{Dots} indicate the peaks of the density distribution of the MCMC chain elements in this diagram, for the different models (color coded as in Table~\ref{t:models}). The \emph{contour} contains 68\% of the MCMC chain elements for model 15e (\emph{navy blue}). We omit the contours of the other models for the sake of clarity.
   The \emph{solid line} is the relation 
   $\alpha_r= \ -{35/ 18}+{2/9}\, \beta_0\ $ from \citet{DML05}.
         \label{f:alpharbeta0}}
   \end{figure}

 \begin{figure}
   \centering
      \includegraphics[width=\hsize]{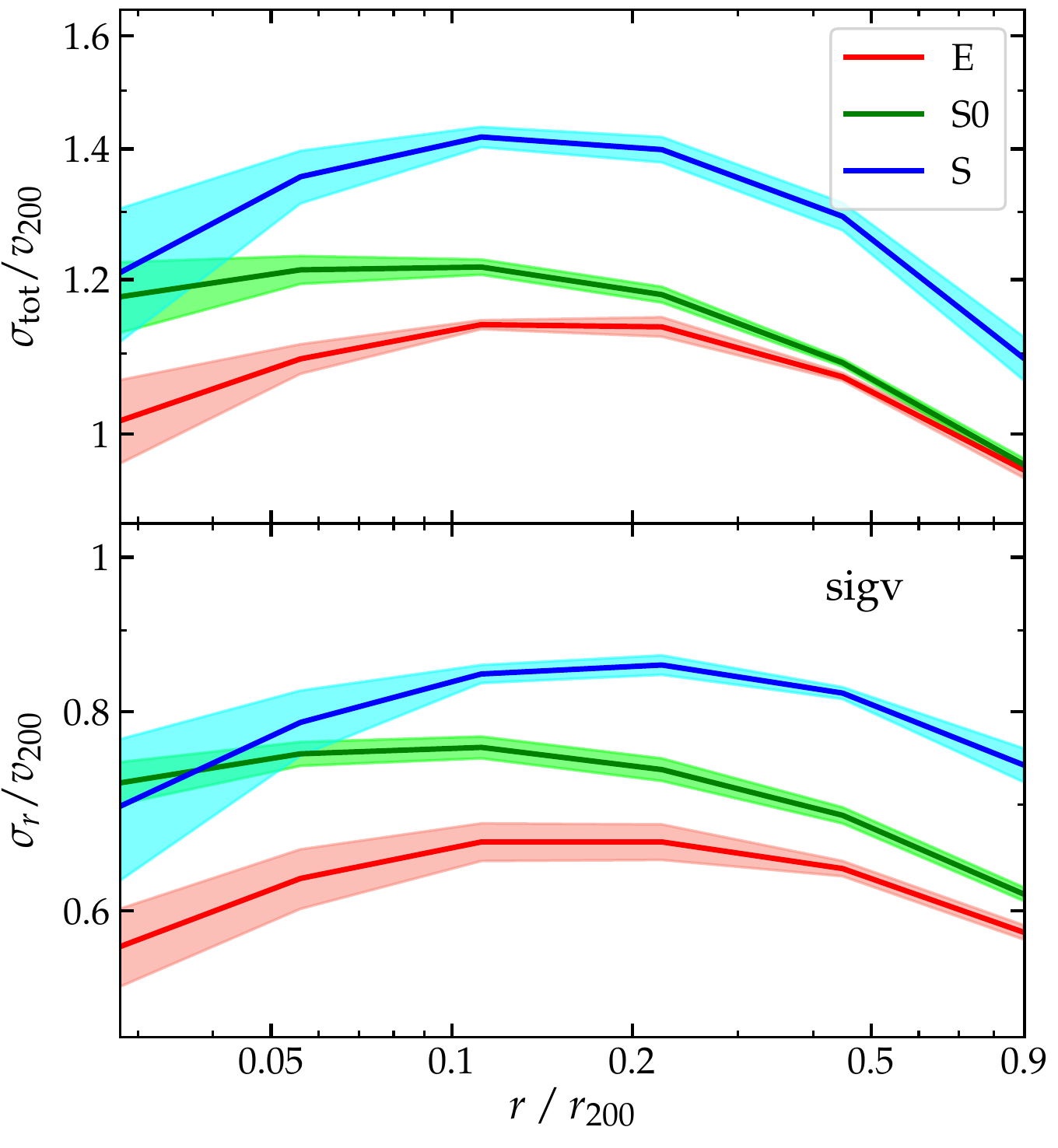}
   \caption{{\em Top panel:} total, $\sigma(r)$, profiles for the ellipticals (\emph{solid red line} and \emph{pink shading}), S0s (\emph{dash-dotted green line} and \emph{light green shading}), and spirals (\emph{dashed blue line} and \emph{cyan shading}). The \emph{curves} are the biweight averages over all 7 models and the shadings are the dispersions among the 7 models. {\em Bottom panel:} same as top panel, but for the radial, $\sigma_r(r)$, profiles. }
         \label{f:vdps}
   \end{figure}

While the $\qnu$ and $\qrnu$ profiles of S0s and spirals differ from the simulation-based expectations for DM particles, it is surprising that their $\qrho$ and $\qrrho$ do not. Then, violent relaxation cannot be the only process conducive to the observed $\qrho$ and $\qrrho$ power-law slopes.
According to \citet{DML05} the dynamical process that leads to the $\qr$ power-law behavior can be understood in terms of the Jeans equation of dynamical equilibrium by assuming that $\beta$ is linearly related to $\gamma$. In their model, the logarithmic slope $\alpha_r$ of $\qr$, must be related to the central orbital anisotropy $\beta_0$ by 
\begin{equation}
\label{e:beta0}
\alpha_r=
\ -{35\over 18}+{2\over 9}\, \beta_0\ .
\end{equation}
In Fig.~\ref{f:alpharbeta0}, we show the distributions of the MCMC chain elements in the $\alpha_r-\beta_0$ plane, separately for the three morphological classes. 
Ellipticals 
follow quite closely \citeauthor{DML05}'s relation  (eq.~[\ref{e:beta0}], above), and so do spirals for most - but not all - models, while S0s do not.
So the dynamical process that leads to the observed $\qrho$ and $\qrrho$ power-law slopes, might indeed be the one suggested by \citeauthor{DML05} for spirals. Even if spirals are only recently accreted to the cluster, and cannot be considered fully dynamically relaxed in the cluster potential, the analysis of semi-analytical simulations indicate that they obey the Jeans equation of dynamical equilibrium \citep{Tagliaferro+21}, so the above interpretation of \citeauthor{DML05} can apply to them.

On the other hand, the process described by \citet{DML05} does not seem to be a viable explanation for the consistency of the $Q_{\rho}(r)$ and $Q_{r,\rho}(r)$ of S0s with simulation-based expectations for DM particles, as they appear to depart from the relation between PPSD slope and inner velocity anisotropy of Eq.~(\ref{e:beta0}). However, among the three morphological classes considered here, S0s show the strongest, albeit not very significant, deviation of the $\qrho$ and $\qrrho$ profile slopes from the simulation-based expectations (see Fig.~\ref{f:qdev}). In \citetalias{Cava+17} we argued that S0s are a transition class between the spiral and elliptical classes, as far as their dynamics within the cluster is concerned. Their velocity dispersion profile appears to be close to that of spirals near the center and to that of ellipticals in the outer regions. This is true not only for the line-of-sight velocity dispersion profile, as we noted in \citetalias{Cava+17} already, but also when considering the total, $\sigma(r)$, and radial, $\sigma_r(r)$, profiles, as shown in Fig.~\ref{f:vdps}. On the other hand, the ellipticals and spirals have very similar $\sigma(r)$ and $\sigma_r(r)$, except for different normalizations, as expected from the similarity of the logarithmic slopes of their $Q_{\rho}$ and $Q_{r,\rho}$ profiles.

It is possible that S0s are not a homogeneous class, but a mixed bag of galaxies that formed in different ways at different epochs of the cluster evolution, namely by ram pressure stripping of disks \citep{GG72} and by merger growth of bulges \citep{vandenBergh90}. The two formation channels of S0s is suggested by studies of their internal structure, gas content, and kinematics \citep{Coccato+20,Deeley+20,Deeley+21}, with disk stripping dominating in clusters and bulge growth in isolated galaxies \citep{Deeley+20}. So maybe the $\qrho$ and $\qrrho$ profiles of S0s agree with simulation-based expectations (albeit less well than those of ellipticals and spirals)
because some S0s followed the dynamical history of ellipticals and some that of spirals. 

We are thus led to suggest the following conclusion. $\qnu(r)$ and $\qrnu(r)$ keep memory of the accretion time of the cluster population, while $\qrho(r)$ and $\qrrho(r)$ are related to the dynamical equilibrium of the population within the cluster potential, that is not necessarily achieved via violent relaxation only.

\subsection{Discussion of results on {\tt num} and {\tt tempX} stacked clusters}
\label{ss:numtempx}

We now turn to the results of our analysis using the other two stacking methods (to determine the virial radii): {\tt num} (richness) and {\tt tempX} (X-ray temperature). The tables and figures are displayed in Appendix~\ref{s:scalings}.

Fig.~\ref{f:linmodnum} and Fig.~\ref{f:linmodtx}
show the linear fractions, $f_l$, of $Q$ and $Q_r$ profiles from the MCMC chain elements for the \texttt{num} and \texttt{tempX} scalings, respectively. One sees $f_l$ values as low or even a bit lower than 40\%, depending on the model and the galaxy type, considerably lower than the $>95\%$ obtained for the \texttt{sigv} scaling. This indicates that the ensemble cluster built using the  \texttt{sigv} scaling has a (projected) phase-space distribution that is more similar to that of simulated halos, than the ensemble clusters built using the other two scalings. 
Another remarkable difference of the \texttt{num} and \texttt{tempX} scalings is that $f_l$ for $Q_r$ profiles is on average lowest for ellipticals among the three morphological classes, while it is lowest for S0s when considering the \texttt{sigv} scaling.

The marginal distributions of the best-fit logarithmic slopes of $Q(r)$ and $\qr(r)$ (considering only linear profiles) are displayed
in Figs.~\ref{f:qdistnum} and \ref{f:qrdistnum} for the \texttt{num} scaling, respectively, 
(left panel: $\qrho$, right panel: $\qnu$) and in Figs.~\ref{f:qdisttx} and \ref{f:qrdisttx} for the \texttt{tempX} scaling. 
For the \texttt{num} scaling, we show in Figs.~\ref{f:qdevnum} 
the averages and dispersions of the $Q(r)$ and $Q_r(r)$ logarithmic slopes obtained on the MCMC chain elements (considering only linear profiles). Fig.~\ref{f:qdevtx}
shows the corresponding quantities for the \texttt{tempX} scaling.  We also provide the average and dispersion of the logarithmic slopes of the $Q$ and $Q_r$ profiles for all MCMC chain elements with linear PPSDs, for all models and all galaxy types in Tables \ref{t:qqrnum} and \ref{t:qqrtx} for the \texttt{num} and \texttt{tempX} scaling, respectively. 

               \begin{figure}
   \centering
\includegraphics[width=\hsize]{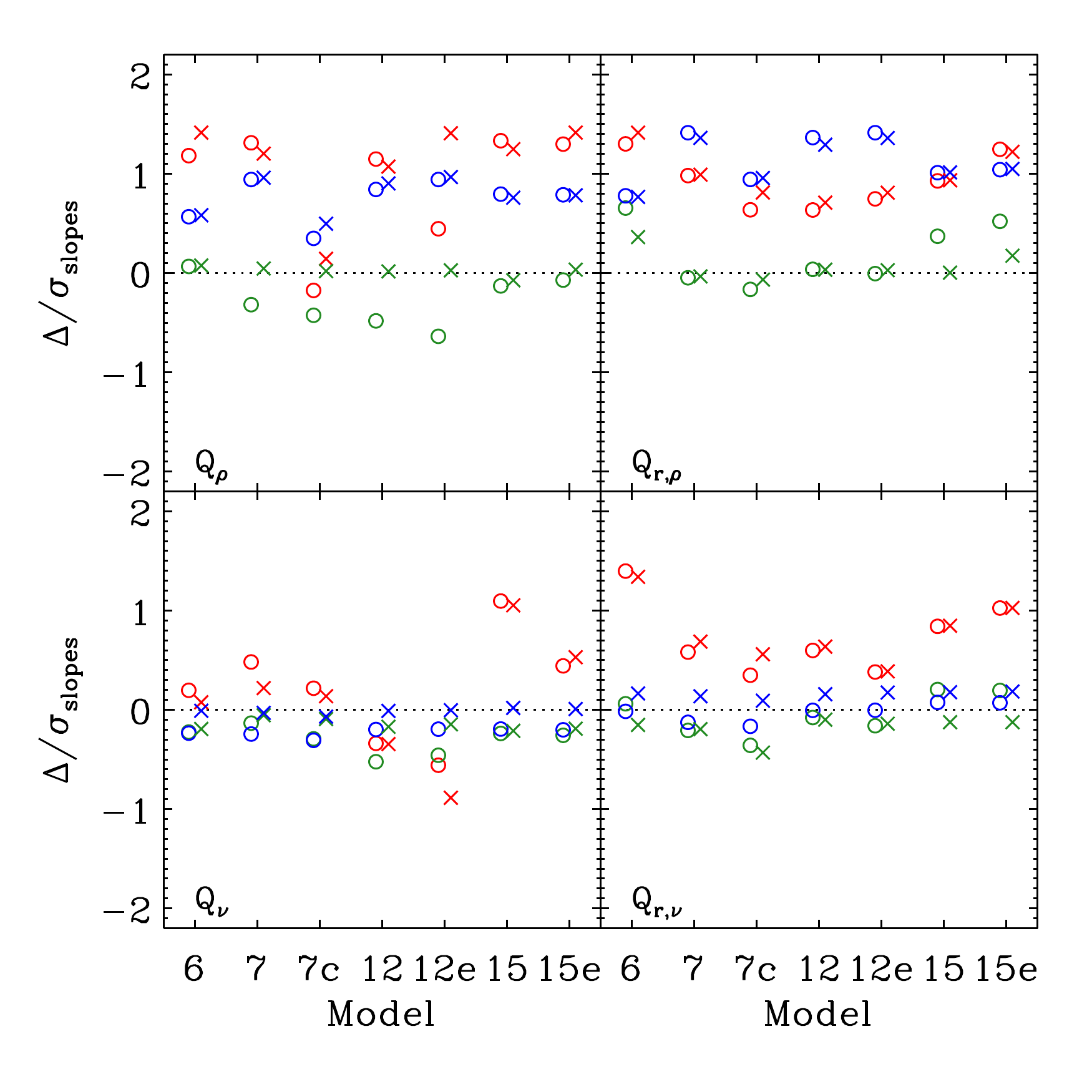}

   \caption{Difference $\Delta$ between the logarithmic slopes obtained for 
   the \texttt{num} (\emph{circles}) and \texttt{tempX} (\emph{crosses}) scalings   and the slopes obtained for the \texttt{sigv} scaling, for the three morphological classes, ellipticals (\emph{red}), S0s (\emph{green}), spirals (\emph{blue}), for the different models ($x$ axis). The $\Delta$ differences are given in units of the quadratically combined dispersions of the slopes.}
         \label{f:qqr_scalings}
         \end{figure}
         
The results for the slopes of $\qrho(r)$ and $\qrrho(r)$ obtained using the \texttt{num} and \texttt{tempX} scalings are generally within one standard deviation of the results obtained using the \texttt{sigv} scaling. This is illustrated in Fig.~\ref{f:qqr_scalings}, where we show the differences $\Delta$ between both the \texttt{num}- and the \texttt{tempX}-scaling slopes and the \texttt{sigv}-scaling slope, considering only linear profiles among all MCMC chains.  The differences are shown in units of the quadratically combined dispersions of the slopes, $\sigma_{{\rm slopes}}$. These differences
 are not statistically significant. 
 The most significant differences come from the $\qr(r)$ slopes of ellipticals and spirals, which are almost identical to that of S0s, and they are all somewhat flatter than the expected relations from numerical simulations (see the top-right panels of Figs.~\ref{f:qdevnum} and \ref{f:qdevtx}).
Moreover, the $Q(r)$ slopes of S0s are intermediate between those of ellipticals and spirals, unlike what was found with the \texttt{sigv} scaling. 

S0s also appear to be intermediate between ellipticals and spirals in the $\beta_0-\alpha_r$ diagram. As seen in Figs.~\ref{f:alpharbeta0num} and \ref{f:alpharbeta0Tx}, it is not the S0s, but the spirals that are the most distant from the expected relation, contrary to what was found using the \texttt{sigv} scaling. Moreover, the velocity dispersion profiles of S0s show less of a transition from those of spirals at small radii to those of ellipticals near the virial radius (Figs.~\ref{f:vdpsnum} and \ref{f:vdpstx}) than is the case for the {\tt sigv} stack (Fig.~\ref{f:vdps}).
The results for the \texttt{num} and \texttt{tempX} scalings therefore suggest that S0s are an intermediate class between ellipticals and spirals, rather than a mixed class. Another remarkable difference with respect to the \texttt{sigv} scaling, is that the $\beta_0-\alpha_r$ relation of Eq.~(\ref{e:beta0}) is not obeyed by any of the three morphological classes. This means we cannot rely on \citet{DML05}'s explanation for why later accreted galaxy populations such as the spirals, and to a lesser extent, S0s, have $Q(r)$ profiles consistent with those of DM particles.

\begin{figure}
    \centering
    \includegraphics[width=\hsize]{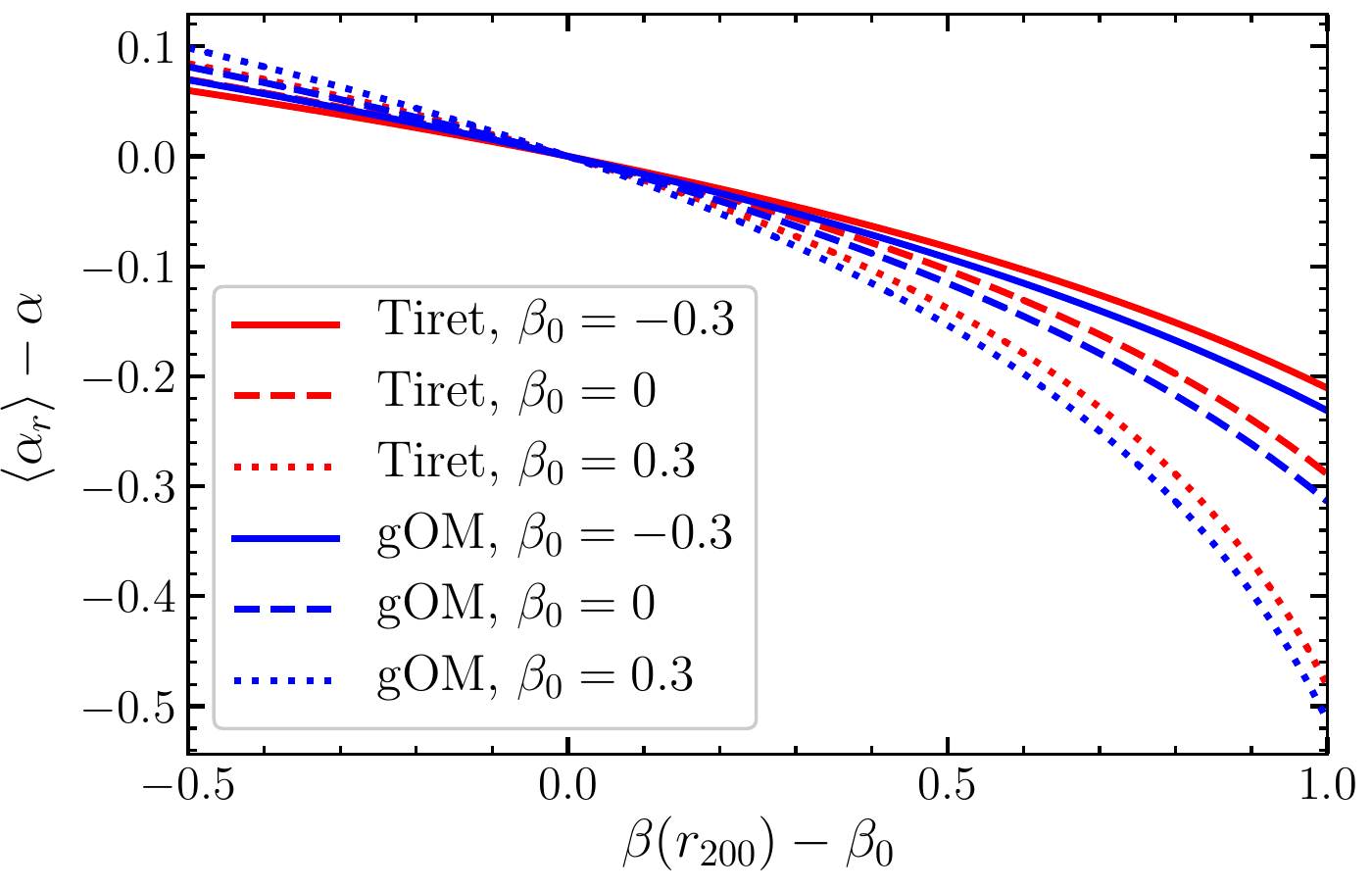}
    \caption{Difference of mean logarithmic slope of $\qrrho$ with logarithmic slope of $\qrho$ (here $\alpha=-1.8$) as a function of difference in velocity anisotropies between virial radius and 0, using Eqs.~(\ref{betaofrmodel}) and (\ref{fofy1}) for $c = \rvir/r_\beta=4$. Changing $c$ and $\alpha$ has negligible effect on the curves. This indicates that $\qrrho$ is steeper than $\qrho$ unless $\beta(\rvir)<\beta_0$.} 
    \label{fig:delalphar}
\end{figure}

Not only are the $\qrrho$ profiles obtained using the \texttt{num} and \texttt{tempX} scalings flatter than simulations predict for DM particles, they are in some cases even flatter than the $\qrho$ profiles. This can happen if
the velocity anisotropy profiles are more radial near the center than at the cluster virial radius, as illustrated in Fig.~\ref{fig:delalphar}. Anisotropy profiles of this kind are not typical of either simulated cluster-size halos \citep[e.g.,][]{Ascasibar&Gottlober08,MBM10,Lemze+12,Munari+13,Lotz+19} or real clusters \citep[e.g.][]{NK96,BK03,Lemze+09,WL10,Biviano+13,Annunziatella+16,AADDV17,Capasso+19}. This suggests that one should take the results obtained using the \texttt{num} and \texttt{tempX} scalings with some caution.

In conclusion, while the results we obtain for the \texttt{num} and \texttt{tempX} scalings are not significantly different from those obtained for the \texttt{sigv} scaling,  they are more distant from the predictions from numerical simulations for what concerns the linearity of the profiles and the slope of $\qrrho(r)$. If the power-law behavior of $Q(r)$ and $\qr(r)$ could be theoretically motivated, the better adherence of the \texttt{sigv}-based profiles to the power-law behavior would suggest that the velocity dispersion is a better $\rvir$ estimator than either the cluster richness or its X-ray temperature, at least for the WINGS cluster data set.

\section{Summary and conclusions} \label{s:conc}
We determined the average $Q$ and $\qr$ profiles of nearby galaxy clusters, using either total mass density $\rho(r)$ or tracer number density $\nu(r)$, as well as the velocity dispersion profiles of three galaxy classes, ellipticals, S0s, and spirals. For this, we have used the results of the MCMC analysis of the kinematics of a velocity-dispersion based (\texttt{sigv}) stack of 54 regular clusters \citepalias{Cava+17} from the WINGS dataset \citep{Fasano+06,Cava+09,Moretti+14} performed with the MAMPOSSt code in \citetalias{Mamon+19}. 

We find that $\qrho(r)$ and $\qrrho(r)$ are very close to the power-law relations predicted by numerical simulations for DM particles \citep{TN01,RTM04,DML05}, at least in a range from a few percent to one virial radius. On the other hand, $\qnu(r)$ and $\qrnu(r)$ agree with the simulation-based predictions for DM particles only for the ellipticals, and deviate marginally and significantly from the simulation-based predictions for the S0s and spirals, respectively. Only the spiral $\qnu(r)$ is  similar to that of subhalos in halos from cosmological hydrodynamical simulations. 

We checked our results on two different stacks of the same data set, based on richness (\texttt{num}) and gas temperature (\texttt{tempX}) scalings. While we find a lower fraction of power-law $Q$ and $\qr$ profiles, the average slopes of these profiles are not significantly different from those obtained for the \texttt{sigv} scaling.

We argue that our results based on the \texttt{sigv} scaling support a scenario in which $\qrho(r)$ and $\qrrho(r)$ are either established early on, during the cluster violent relaxation phase, for the DM and ellipticals, or established subsequently,
for spirals by adapting their orbital and spatial distribution as they move towards dynamical equilibrium in the cluster potential. 
S0s might be a mixed class, part of them following the dynamical history of ellipticals, and the other part, that of spirals, as suggested by our analysis of the {\tt sigv} stack, or an intermediate class between spirals and ellipticals as consistent with our analysis of the {\tt num} and {\tt tempX} stacks. $\qnu(r)$ and $\qrnu(r)$ are not universal, and depend on the time of accretion of the tracer population in the cluster. 

In conclusion, our results give strong observational support to the simulation-based power-law $Q$ and $\qr$ profiles when they are defined using total mass density $\rho(r)$ rather than the tracer number density $\nu(r)$. 

\begin{acknowledgements}
      We thank the referee for her or his constructive and pertinent comments. We also acknowledge the WINGS team for their active and precious collaboration. We thank Ilaria Marini for useful discussions. GAM and AB are grateful to the IFPU and IAP, respectively for their hospitality during part of this collaboration.
\end{acknowledgements}

\bibliography{master}

\begin{appendix}

\section{Results for the \texttt{num} and \texttt{tempX} scalings}
\label{s:scalings}
In the main text of this paper, we provided the results for the velocity dispersion-based \texttt{sigv} scaling used to stack the clusters.  Here we provide the results for the richness-based, \texttt{num}, and X-ray temperature-based, \texttt{tempX} scalings. 

    \begin{figure}
   \centering
   \includegraphics[width=\hsize]{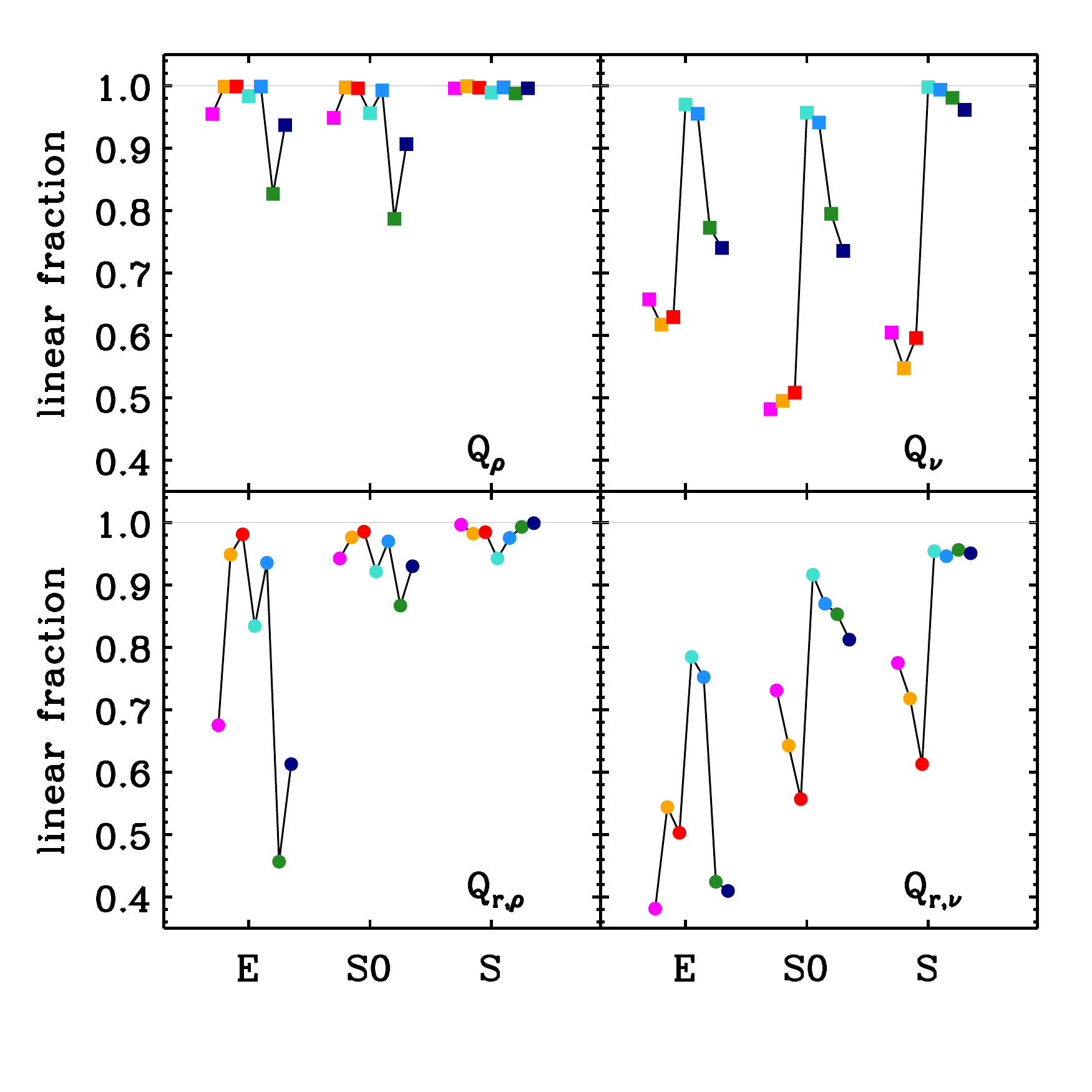}
      \caption{Same as Fig.~\ref{f:linmod} but for the \texttt{num} scaling.
       } 
         \label{f:linmodnum}
   \end{figure}
    \begin{figure}
   \centering
   \includegraphics[width=\hsize]{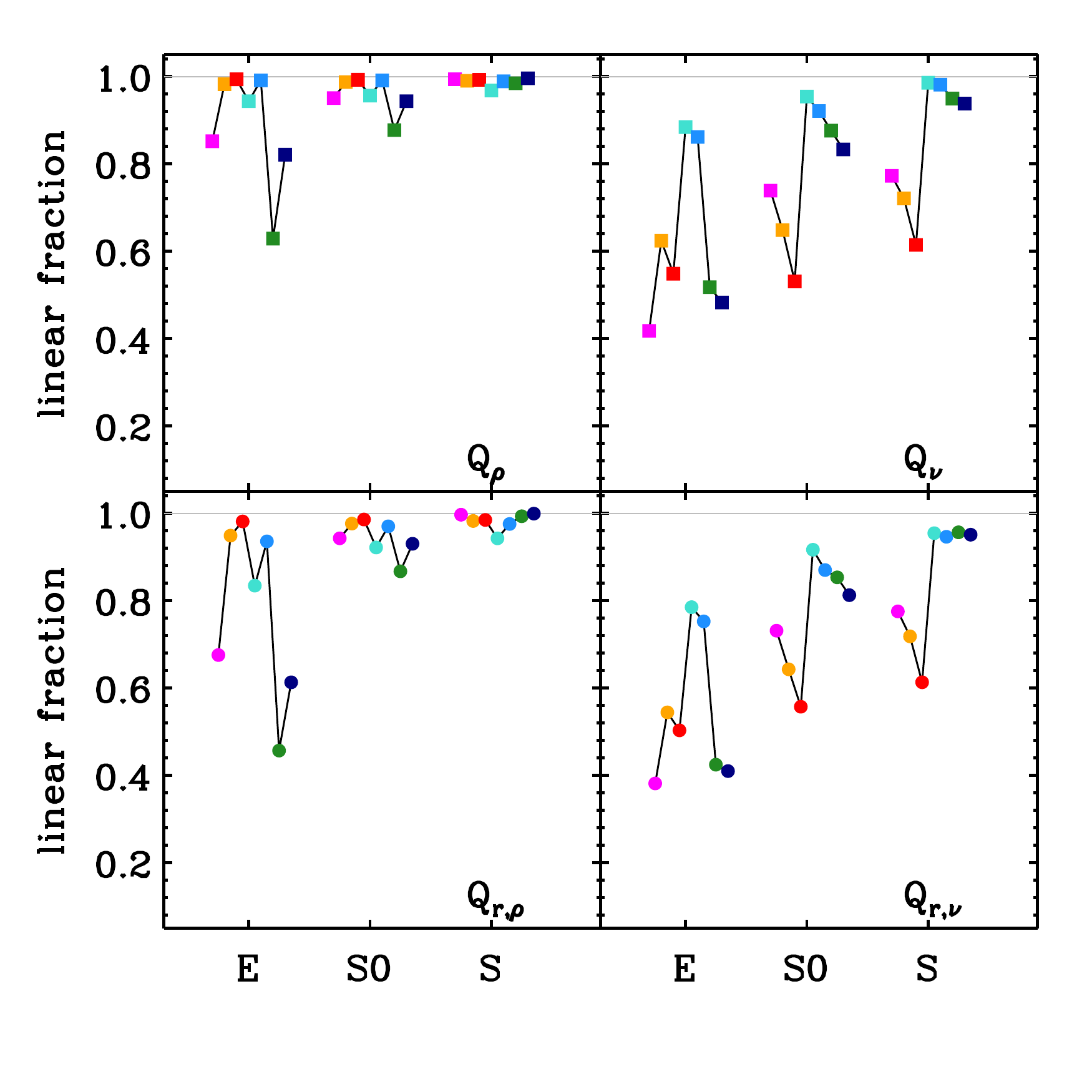}
      \caption{Same as Fig.~\ref{f:linmod} but for the \texttt{tempX} scaling. } 
         \label{f:linmodtx}
   \end{figure}

\begin{table*}
\centering
\caption{$Q$ and $Q_r$ profiles: $f_l$ and slopes for the \texttt{num} scaling}
\label{t:qqrnum}
\resizebox{\textwidth}{!}{
\begin{tabular}{r|cc|cc|cc||cc|cc|cc}
  \hline \\
   & \multicolumn{6}{c}{$\qrho$} & \multicolumn{6}{c}{$\qnu$} \\ \\
        & \multicolumn{2}{c}{E} & \multicolumn{2}{c}{S0} & \multicolumn{2}{c}{S} & \multicolumn{2}{c}{E} & \multicolumn{2}{c}{S0} & \multicolumn{2}{c}{S} \\
        Model &  $f_l$ & slope & $f_l$ & slope & $f_l$ & slope &  $f_l$ & slope & $f_l$ & slope & $f_l$ & slope \\
        \hline
6    &  0.95  & $-1.83 \pm  0.08 $ &  0.95  & $-1.75 \pm  0.10 $ &  1.00  & $-1.70 \pm  0.12 $ &  0.66  & $-1.74 \pm  0.07 $ &  0.48  & $-1.67 \pm  0.07 $ &  0.60  & $-1.42 \pm  0.09$ \\
 7    &  1.00  & $-1.82 \pm  0.07 $ &  1.00  & $-1.77 \pm  0.08 $ &  1.00  & $-1.72 \pm  0.13 $ &  0.62  & $-1.72 \pm  0.07 $ &  0.50  & $-1.67 \pm  0.07 $ &  0.55  & $-1.42 \pm  0.09$ \\
 7c   &  1.00  & $-1.82 \pm  0.05 $ &  1.00  & $-1.77 \pm  0.05 $ &  1.00  & $-1.66 \pm  0.08 $ &  0.63  & $-1.76 \pm  0.07 $ &  0.51  & $-1.71 \pm  0.07 $ &  0.60  & $-1.48 \pm  0.08$ \\
12    &  0.98  & $-1.82 \pm  0.08 $ &  0.96  & $-1.79 \pm  0.09 $ &  0.99  & $-1.73 \pm  0.14 $ &  0.97  & $-1.86 \pm  0.10 $ &  0.96  & $-1.77 \pm  0.10 $ &  1.00  & $-1.37 \pm  0.10$ \\
12e   &  1.00  & $-1.85 \pm  0.07 $ &  0.99  & $-1.81 \pm  0.09 $ &  1.00  & $-1.75 \pm  0.14 $ &  0.96  & $-1.83 \pm  0.09 $ &  0.94  & $-1.74 \pm  0.09 $ &  0.99  & $-1.37 \pm  0.10$ \\
15    &  0.83  & $-1.80 \pm  0.07 $ &  0.79  & $-1.75 \pm  0.09 $ &  0.99  & $-1.72 \pm  0.14 $ &  0.77  & $-1.82 \pm  0.10 $ &  0.79  & $-1.71 \pm  0.10 $ &  0.98  & $-1.37 \pm  0.10$ \\
15e   &  0.94  & $-1.82 \pm  0.08 $ &  0.91  & $-1.76 \pm  0.09 $ &  1.00  & $-1.73 \pm  0.13 $ &  0.74  & $-1.81 \pm  0.10 $ &  0.74  & $-1.71 \pm  0.09 $ &  0.96  & $-1.37 \pm  0.09$ \\
 & & & & & & & & & & \\
mean & 0.96 & $-1.82 \pm 0.01$ & 0.94 & $-1.77 \pm 0.02$ & 0.99 & $-1.71 \pm 0.03$ & 0.76 & $-1.78 \pm 0.05$ & 0.70 & $-1.71 \pm 0.04$ & 0.81 & $-1.40 \pm 0.04$ \\
\hline \\
   & \multicolumn{6}{c}{$\qrrho$} & \multicolumn{6}{c}{$\qrnu$} \\ \\
        & \multicolumn{2}{c}{E} & \multicolumn{2}{c}{S0} & \multicolumn{2}{c}{S} & \multicolumn{2}{c}{E} & \multicolumn{2}{c}{S0} & \multicolumn{2}{c}{S} \\
  Model      &  $f_l$ & slope & $f_l$ & slope & $f_l$ & slope &  $f_l$ & slope & $f_l$ & slope & $f_l$ & slope \\
\hline
6    &  0.91  & $-1.85 \pm  0.24 $ &  0.93  & $-1.71 \pm  0.21 $ &  1.00  & $-1.71 \pm  0.17 $ &  0.64  & $-1.84 \pm  0.18 $ &  0.47  & $-1.71 \pm  0.16 $ &  0.61  & $-1.49 \pm  0.15$ \\
 7    &  0.99  & $-1.84 \pm  0.20 $ &  0.99  & $-1.78 \pm  0.19 $ &  1.00  & $-1.80 \pm  0.27 $ &  0.55  & $-1.81 \pm  0.17 $ &  0.47  & $-1.75 \pm  0.15 $ &  0.62  & $-1.53 \pm  0.15$ \\
 7c   &  1.00  & $-1.85 \pm  0.14 $ &  0.99  & $-1.78 \pm  0.14 $ &  1.00  & $-1.72 \pm  0.19 $ &  0.58  & $-1.87 \pm  0.16 $ &  0.50  & $-1.79 \pm  0.16 $ &  0.64  & $-1.56 \pm  0.14$ \\
12    &  0.91  & $-1.74 \pm  0.16 $ &  0.89  & $-1.72 \pm  0.17 $ &  0.97  & $-1.80 \pm  0.24 $ &  0.89  & $-1.78 \pm  0.19 $ &  0.89  & $-1.71 \pm  0.19 $ &  0.98  & $-1.41 \pm  0.14$ \\
12e   &  0.97  & $-1.77 \pm  0.17 $ &  0.96  & $-1.75 \pm  0.18 $ &  0.99  & $-1.81 \pm  0.24 $ &  0.85  & $-1.79 \pm  0.18 $ &  0.85  & $-1.72 \pm  0.17 $ &  0.97  & $-1.43 \pm  0.15$ \\
15    &  0.71  & $-1.69 \pm  0.16 $ &  0.75  & $-1.63 \pm  0.17 $ &  1.00  & $-1.73 \pm  0.19 $ &  0.69  & $-1.71 \pm  0.19 $ &  0.71  & $-1.60 \pm  0.17 $ &  0.97  & $-1.39 \pm  0.15$ \\
15e   &  0.83  & $-1.71 \pm  0.17 $ &  0.87  & $-1.65 \pm  0.18 $ &  1.00  & $-1.74 \pm  0.19 $ &  0.66  & $-1.74 \pm  0.18 $ &  0.66  & $-1.64 \pm  0.16 $ &  0.95  & $-1.40 \pm  0.14$ \\
 & & & & & & & & & & \\
mean & 0.90 & $-1.77 \pm 0.07$ & 0.91 & $-1.72 \pm 0.06$ & 0.99 & $-1.75 \pm 0.04$ & 0.70 & $-1.79 \pm 0.05$ & 0.65 & $-1.71 \pm 0.06$ & 0.82 & $-1.46 \pm 0.07$ \\
\hline
\end{tabular}}
\tablefoot{Columns labelled `$f_l$' give the fraction of linear MCMC $Q$ profiles. Columns labelled 'slope' give the average and dispersion of the slopes of the MCMC $Q$ profiles with $f_l>0.1$. Rows labelled `mean' give the weighted mean and dispersion of all the models, using the slope dispersions as weights.}
\end{table*}

\begin{table*}
\centering
\caption{$Q$ and $Q_r$ profiles: $f_l$ and slopes for the \texttt{tempX} scaling}
\label{t:qqrtx}
\resizebox{\textwidth}{!}{
\begin{tabular}{r|cc|cc|cc||cc|cc|cc}
  \hline \\
   & \multicolumn{6}{c}{$\qrho$} & \multicolumn{6}{c}{$\qnu$} \\ \\
        & \multicolumn{2}{c}{E} & \multicolumn{2}{c}{S0} & \multicolumn{2}{c}{S} & \multicolumn{2}{c}{E} & \multicolumn{2}{c}{S0} & \multicolumn{2}{c}{S} \\
        Model &  $f_l$ & slope & $f_l$ & slope & $f_l$ & slope &  $f_l$ & slope & $f_l$ & slope & $f_l$ & slope \\
        \hline
6    &  0.85  & $-1.78 \pm  0.09 $ &  0.95  & $-1.75 \pm  0.11 $ &  0.99  & $-1.69 \pm  0.16 $ &  0.42  & $-1.76 \pm  0.10 $ &  0.74  & $-1.66 \pm  0.10 $ &  0.77  & $-1.25 \pm  0.12$ \\
 7    &  0.98  & $-1.81 \pm  0.08 $ &  0.99  & $-1.72 \pm  0.10 $ &  0.99  & $-1.66 \pm  0.15 $ &  0.62  & $-1.76 \pm  0.10 $ &  0.65  & $-1.65 \pm  0.09 $ &  0.72  & $-1.27 \pm  0.11$ \\
 7c   &  0.99  & $-1.81 \pm  0.06 $ &  0.99  & $-1.70 \pm  0.08 $ &  0.99  & $-1.61 \pm  0.11 $ &  0.55  & $-1.77 \pm  0.09 $ &  0.53  & $-1.67 \pm  0.09 $ &  0.61  & $-1.33 \pm  0.11$ \\
12    &  0.94  & $-1.81 \pm  0.08 $ &  0.96  & $-1.72 \pm  0.10 $ &  0.97  & $-1.67 \pm  0.17 $ &  0.88  & $-1.86 \pm  0.12 $ &  0.95  & $-1.71 \pm  0.11 $ &  0.99  & $-1.22 \pm  0.11$ \\
12e   &  0.99  & $-1.83 \pm  0.08 $ &  0.99  & $-1.73 \pm  0.10 $ &  0.99  & $-1.68 \pm  0.16 $ &  0.86  & $-1.84 \pm  0.11 $ &  0.92  & $-1.68 \pm  0.10 $ &  0.98  & $-1.23 \pm  0.11$ \\
15    &  0.63  & $-1.78 \pm  0.08 $ &  0.88  & $-1.74 \pm  0.12 $ &  0.98  & $-1.74 \pm  0.18 $ &  0.52  & $-1.81 \pm  0.11 $ &  0.88  & $-1.71 \pm  0.12 $ &  0.95  & $-1.20 \pm  0.12$ \\
15e   &  0.82  & $-1.79 \pm  0.08 $ &  0.94  & $-1.75 \pm  0.12 $ &  1.00  & $-1.73 \pm  0.17 $ &  0.48  & $-1.80 \pm  0.11 $ &  0.83  & $-1.69 \pm  0.11 $ &  0.94  & $-1.21 \pm  0.11$ \\
 & & & & & & & & & & \\
mean & 0.89 & $-1.80 \pm 0.02$ & 0.96 & $-1.73 \pm 0.02$ & 0.99 & $-1.68 \pm 0.05$ & 0.62 & $-1.80 \pm 0.04$ & 0.78 & $-1.68 \pm 0.02$ & 0.85 & $-1.24 \pm 0.04$ \\
\hline \\
   & \multicolumn{6}{c}{$\qrrho$} & \multicolumn{6}{c}{$\qrnu$} \\ \\
        & \multicolumn{2}{c}{E} & \multicolumn{2}{c}{S0} & \multicolumn{2}{c}{S} & \multicolumn{2}{c}{E} & \multicolumn{2}{c}{S0} & \multicolumn{2}{c}{S} \\
  Model      &  $f_l$ & slope & $f_l$ & slope & $f_l$ & slope &  $f_l$ & slope & $f_l$ & slope & $f_l$ & slope \\
\hline
 6    &  0.68  & $-1.72 \pm  0.24 $ &  0.94  & $-1.81 \pm  0.23 $ &  1.00  & $-1.72 \pm  0.20 $ &  0.38  & $-1.78 \pm  0.20 $ &  0.73  & $-1.76 \pm  0.19 $ &  0.78  & $-1.36 \pm  0.16$ \\
 7    &  0.95  & $-1.76 \pm  0.19 $ &  0.98  & $-1.74 \pm  0.21 $ &  0.98  & $-1.71 \pm  0.25 $ &  0.54  & $-1.77 \pm  0.18 $ &  0.64  & $-1.72 \pm  0.19 $ &  0.72  & $-1.35 \pm  0.16$ \\
 7c   &  0.98  & $-1.80 \pm  0.15 $ &  0.99  & $-1.74 \pm  0.17 $ &  0.98  & $-1.63 \pm  0.19 $ &  0.50  & $-1.82 \pm  0.16 $ &  0.56  & $-1.78 \pm  0.19 $ &  0.61  & $-1.40 \pm  0.15$ \\
12    &  0.83  & $-1.70 \pm  0.18 $ &  0.92  & $-1.71 \pm  0.18 $ &  0.94  & $-1.71 \pm  0.24 $ &  0.78  & $-1.76 \pm  0.23 $ &  0.92  & $-1.70 \pm  0.18 $ &  0.95  & $-1.27 \pm  0.15$ \\
12e   &  0.94  & $-1.73 \pm  0.18 $ &  0.97  & $-1.73 \pm  0.18 $ &  0.98  & $-1.72 \pm  0.24 $ &  0.75  & $-1.78 \pm  0.19 $ &  0.87  & $-1.69 \pm  0.17 $ &  0.95  & $-1.28 \pm  0.15$ \\
15    &  0.46  & $-1.65 \pm  0.16 $ &  0.87  & $-1.74 \pm  0.21 $ &  0.99  & $-1.78 \pm  0.23 $ &  0.42  & $-1.68 \pm  0.20 $ &  0.85  & $-1.71 \pm  0.21 $ &  0.96  & $-1.30 \pm  0.16$ \\
15e   &  0.61  & $-1.67 \pm  0.18 $ &  0.93  & $-1.76 \pm  0.22 $ &  1.00  & $-1.77 \pm  0.21 $ &  0.41  & $-1.72 \pm  0.19 $ &  0.81  & $-1.73 \pm  0.20 $ &  0.95  & $-1.31 \pm  0.16$ \\
 & & & & & & & & & & \\
mean & 0.78 & $-1.72 \pm 0.06$ & 0.94 & $-1.74 \pm 0.03$ & 0.98 & $-1.72 \pm 0.05$ & 0.54 & $-1.76 \pm 0.04$ & 0.77 & $-1.73 \pm 0.03$ & 0.84 & $-1.32 \pm 0.05$ \\
\hline
\end{tabular}}
\tablefoot{Columns labelled '$f_l$' give the fraction of linear MCMC $Q$ profiles. Columns labelled 'slope' give the average and dispersion of the slopes of the MCMC $Q$ profiles with $f_l>0.1$. Rows labelled "mean" gives the weighted mean and dispersion of all the models, using the slope dispersions as weights.}
\end{table*}

       \begin{figure}
   \centering
   \includegraphics[width=\hsize]{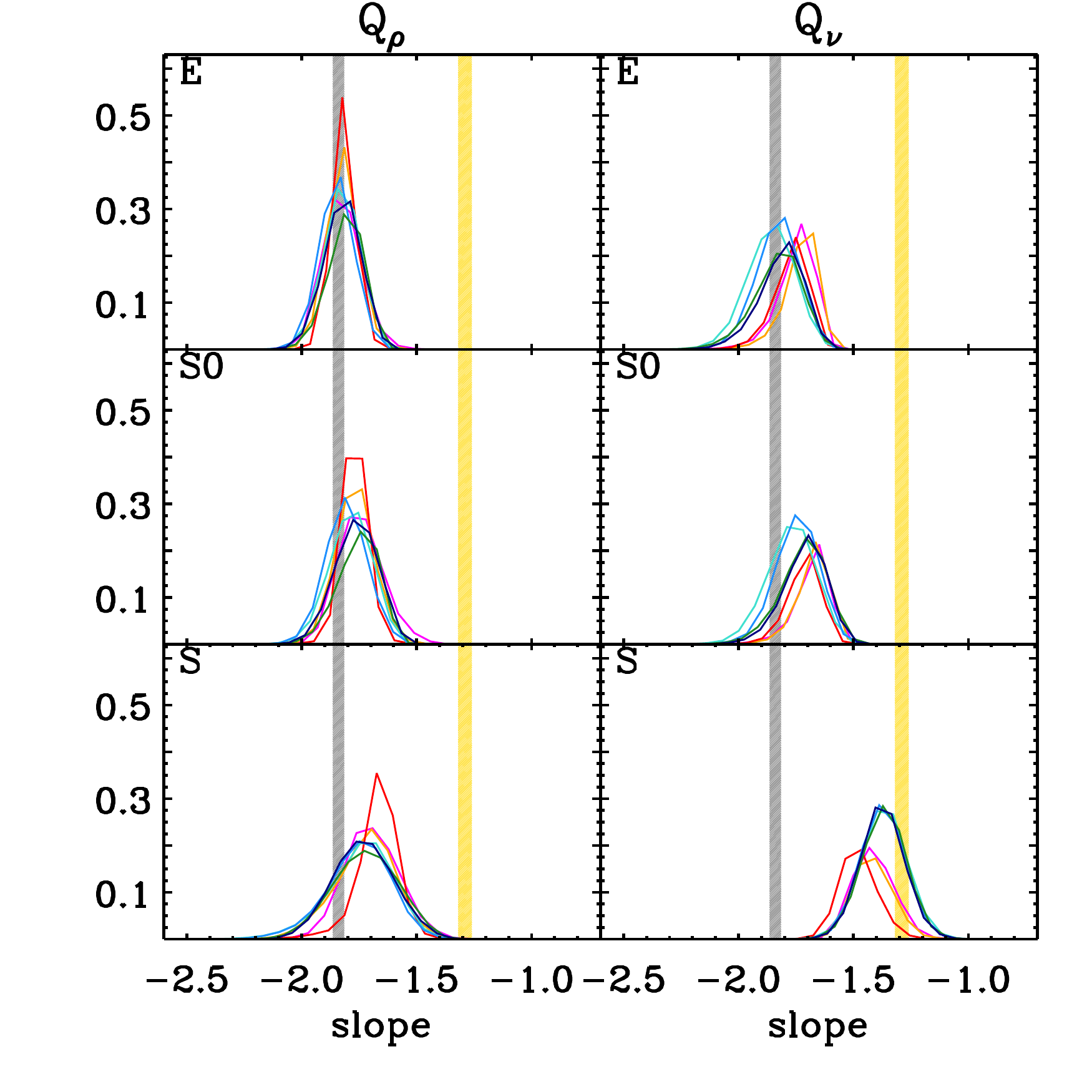}
      \caption{Same as Fig.~\ref{f:qdist} but for the \texttt{num} scaling.} 
         \label{f:qdistnum}
   \end{figure}

       \begin{figure}
   \centering
   \includegraphics[width=\hsize]{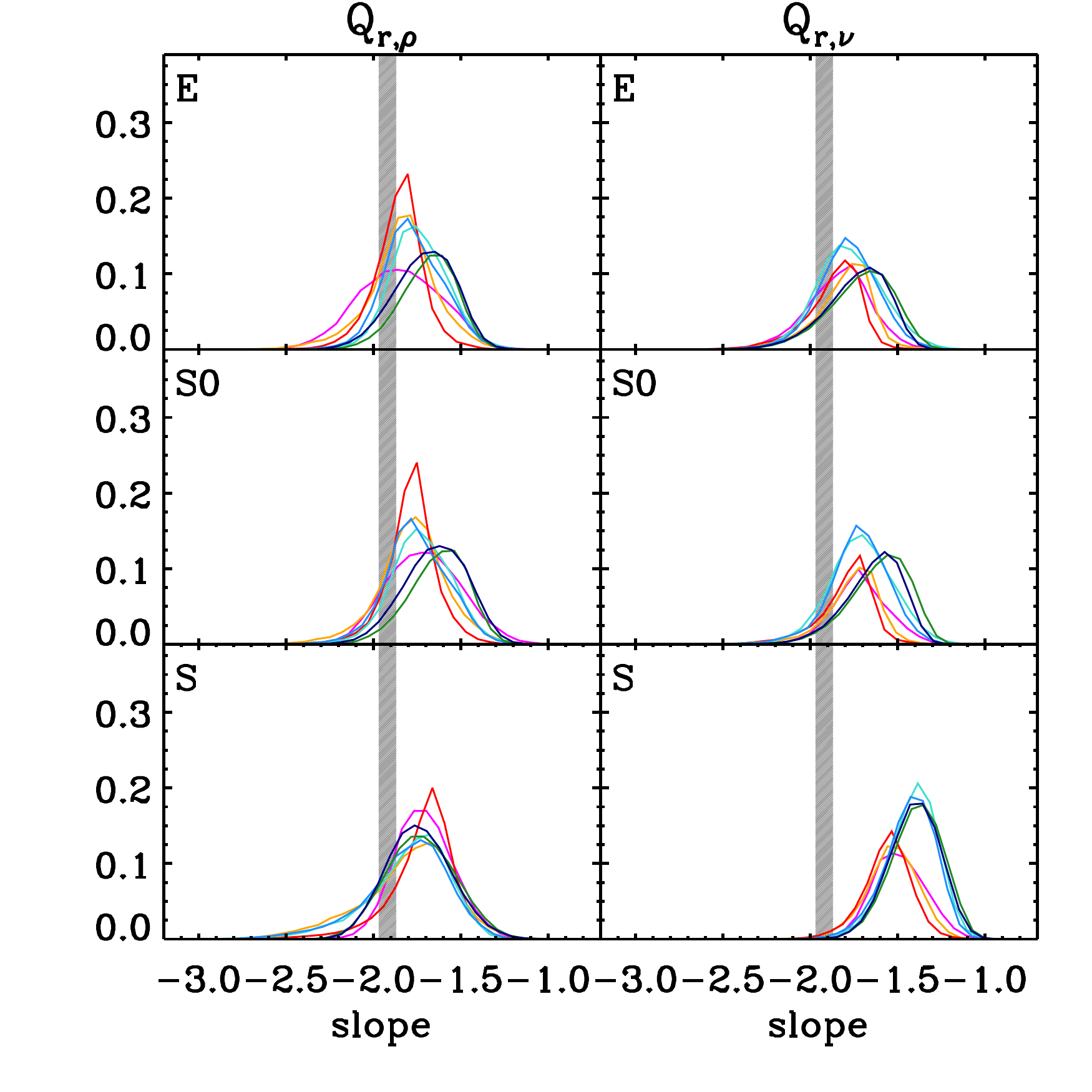}
      \caption{Same as Fig.~\ref{f:qrdist} but for the \texttt{num} scaling.} 
         \label{f:qrdistnum}
   \end{figure}

       \begin{figure}
   \centering
   \includegraphics[width=\hsize]{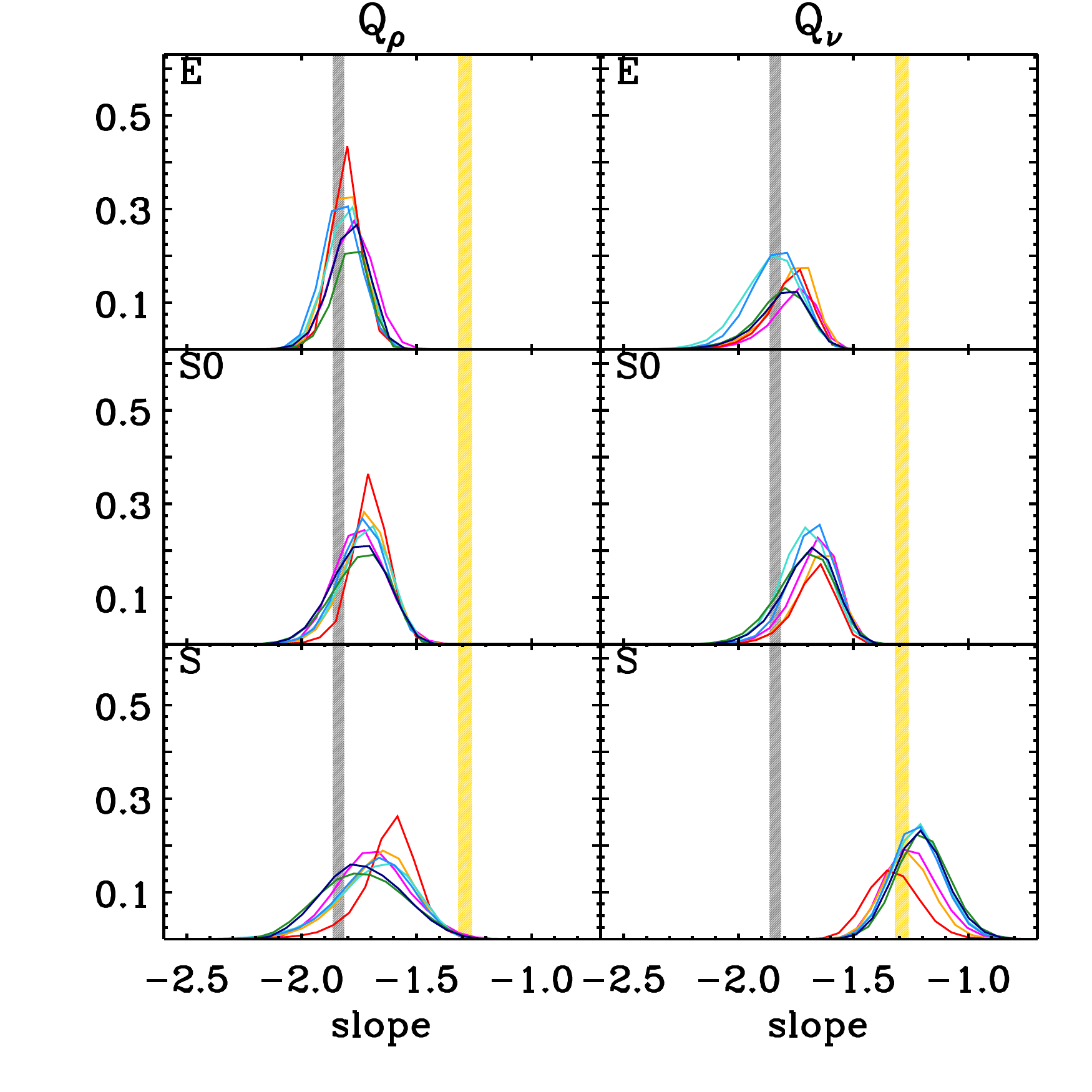}
      \caption{Same as Fig.~\ref{f:qdist} but for the \texttt{tempX} scaling. } 
         \label{f:qdisttx}
   \end{figure}

       \begin{figure}
   \centering
   \includegraphics[width=\hsize]{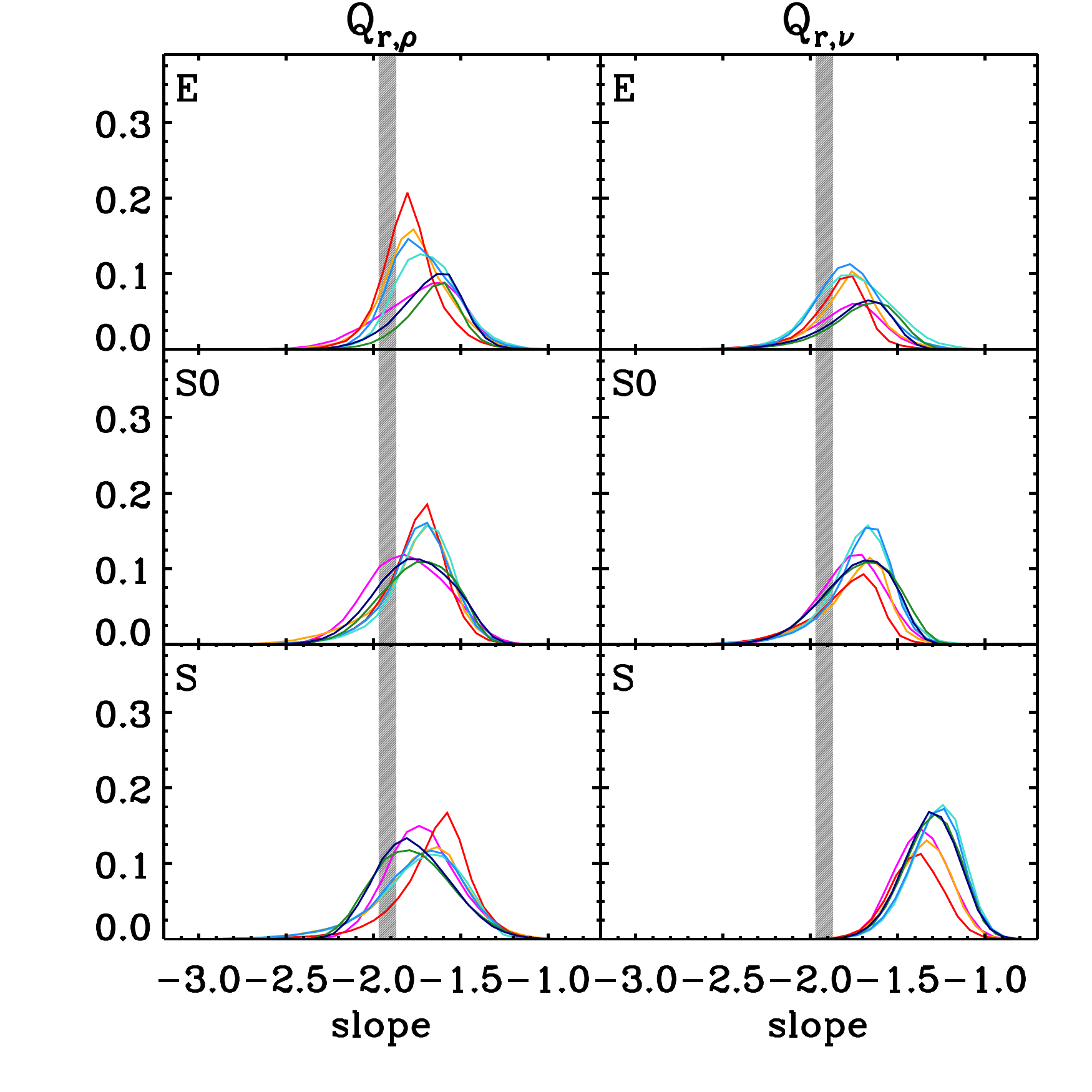}
      \caption{Same as Fig.~\ref{f:qrdist} but for the \texttt{tempX} scaling. } 
         \label{f:qrdisttx}
   \end{figure}

    \begin{figure}
   \centering
   \includegraphics[width=\hsize]{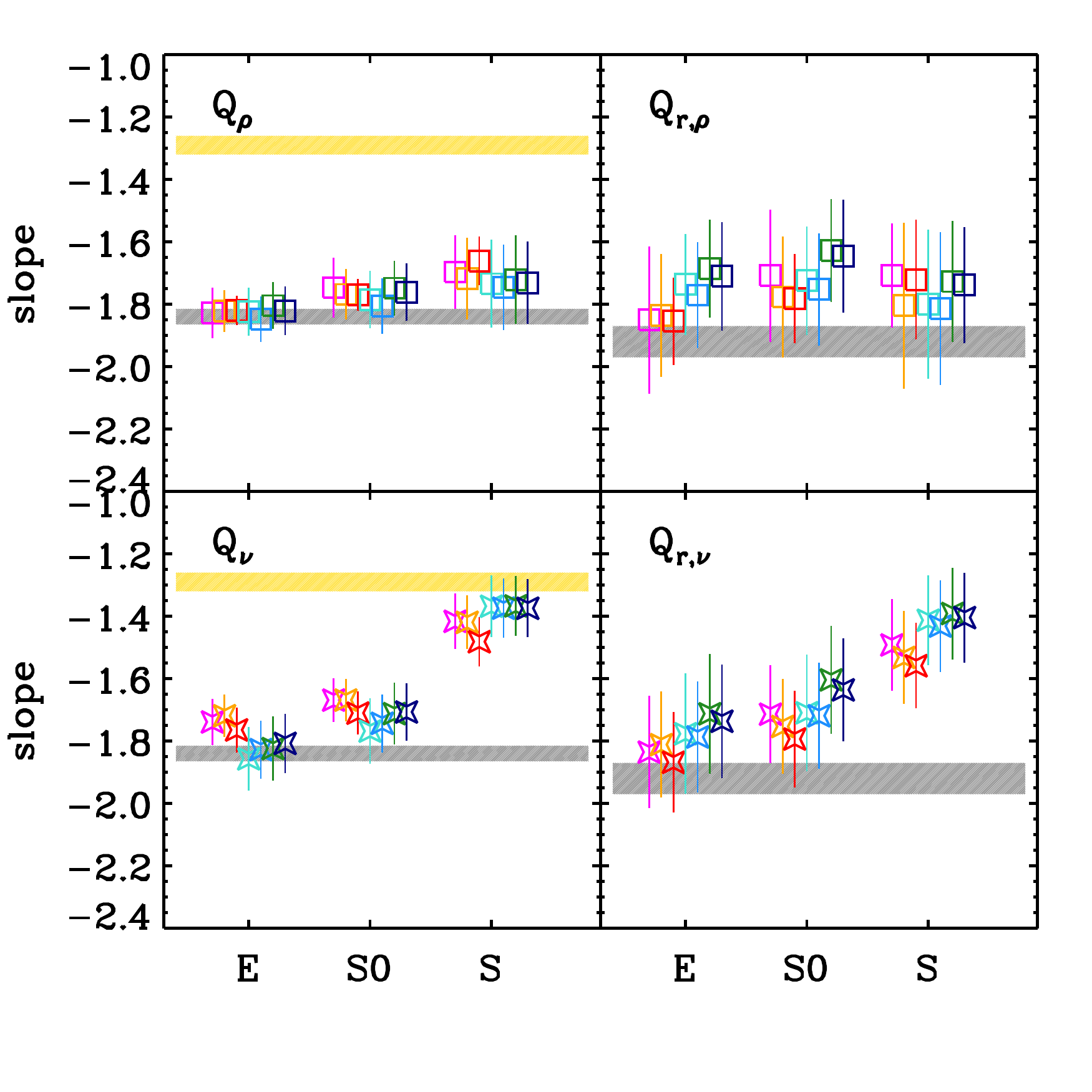}
   \caption{Same as Fig.~\ref{f:qdev} but for the \texttt{num} scaling.}
         \label{f:qdevnum}
         \end{figure}

             \begin{figure}
   \centering
   \includegraphics[width=\hsize]{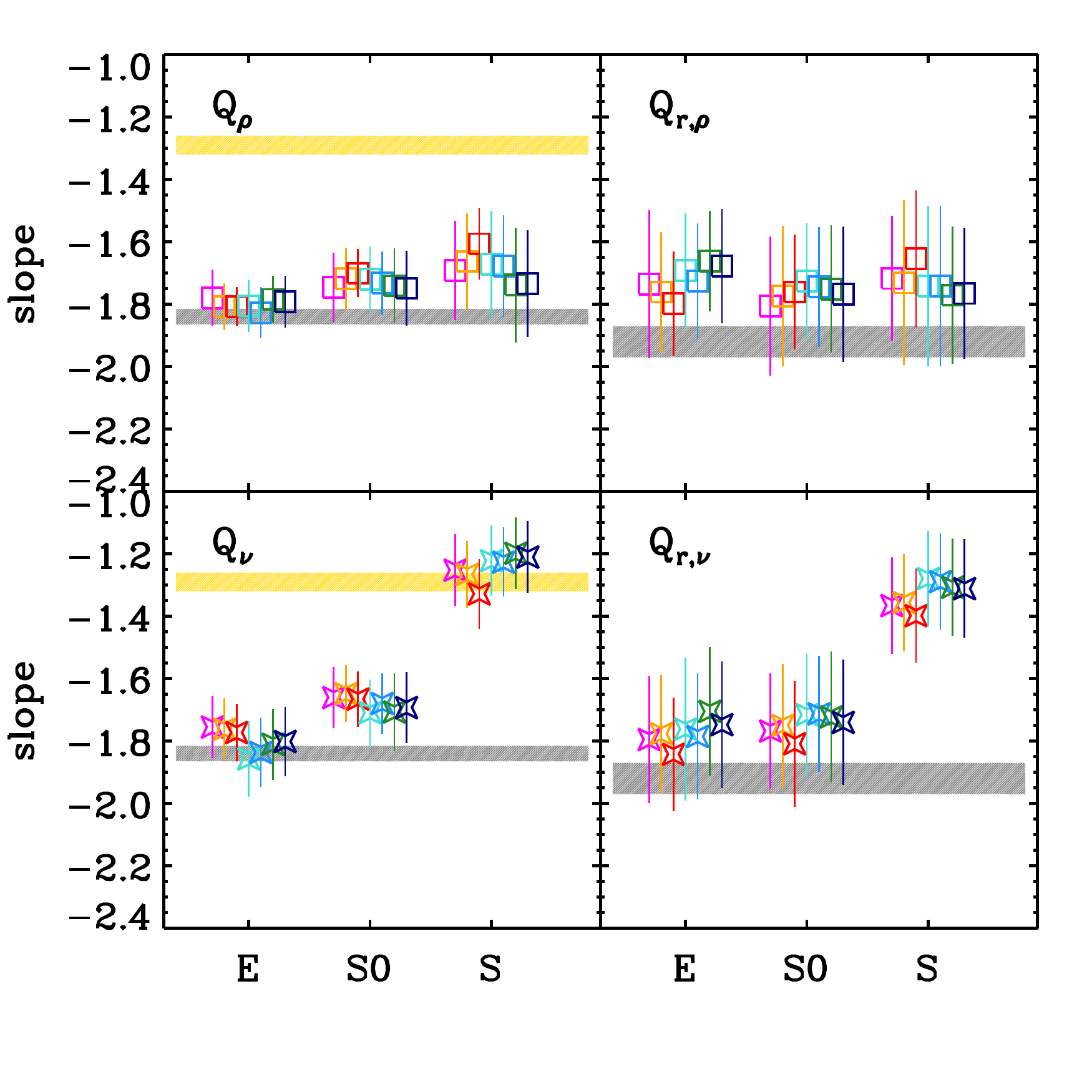}
   \caption{Same as Fig.~\ref{f:qdev} but for the \texttt{tempX} scaling.}
         \label{f:qdevtx}
         \end{figure}

         \begin{figure}
   \centering
   \includegraphics[width=\hsize]{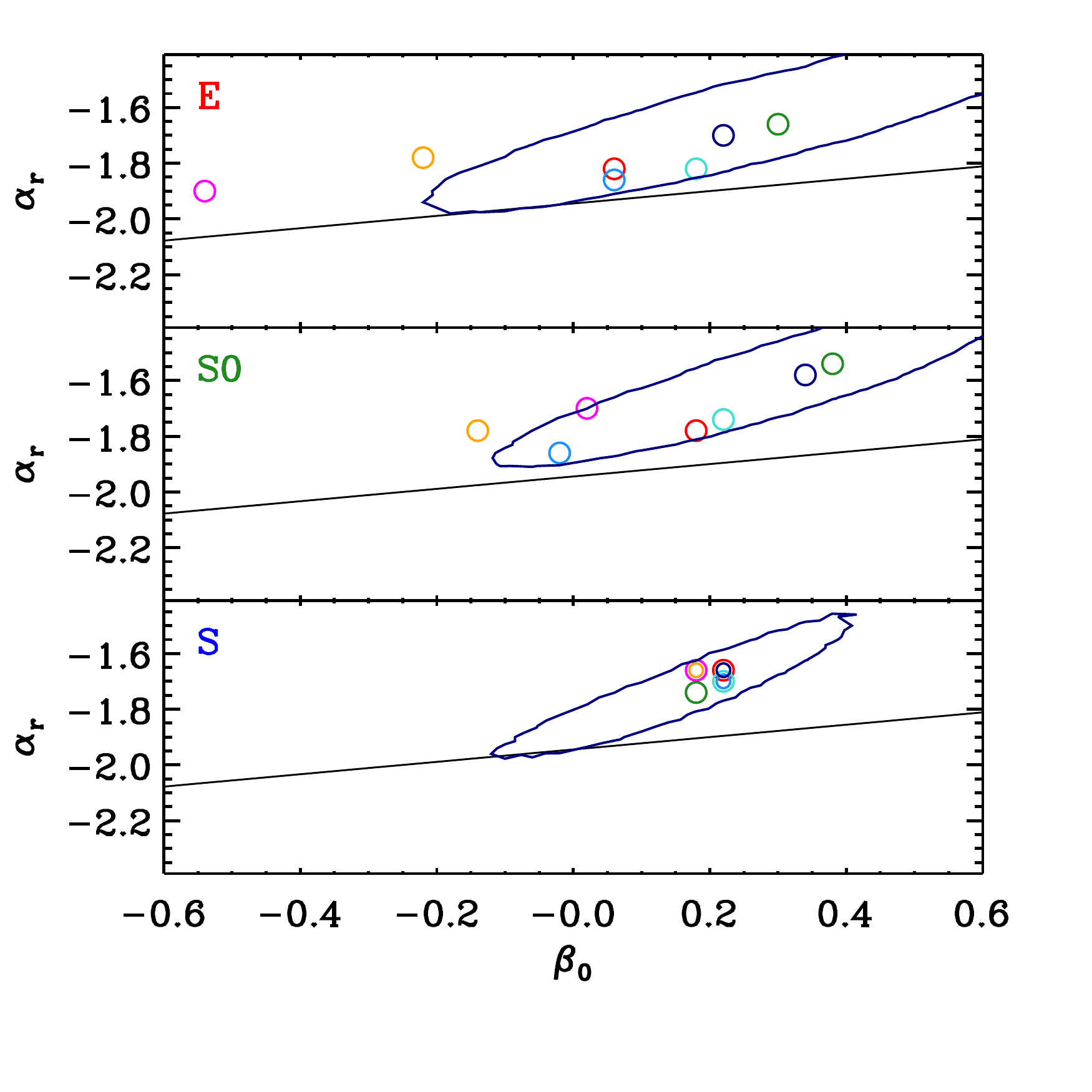}
   \caption{Same as Fig.~\ref{f:alpharbeta0} but for the \texttt{num} scaling.}
         \label{f:alpharbeta0num}
         \end{figure}
           \begin{figure}
   \centering
   \includegraphics[width=\hsize]{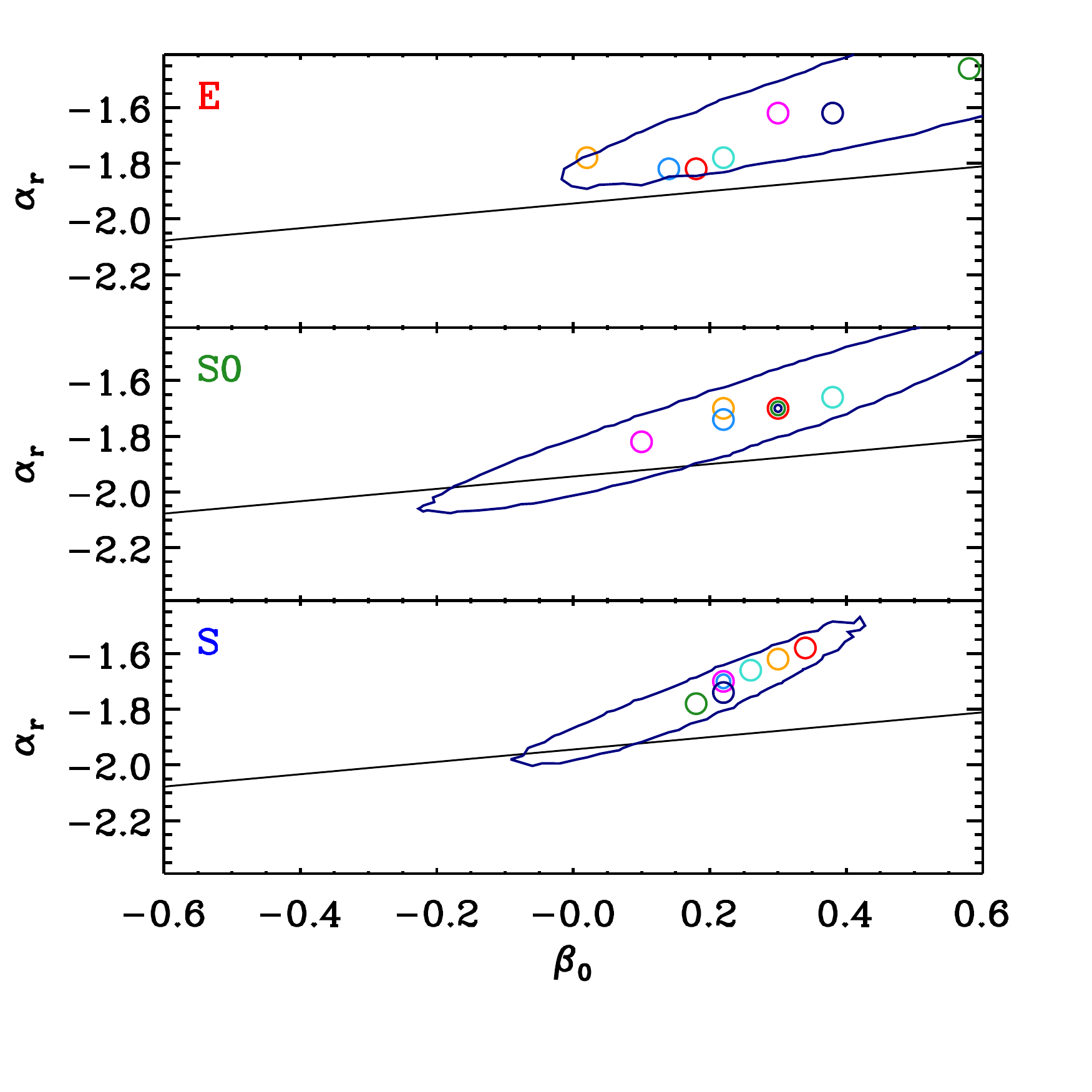}
   \caption{Same as Fig.~\ref{f:alpharbeta0} but for the \texttt{tempX} scaling.}
         \label{f:alpharbeta0Tx}
         \end{figure}
    
     \begin{figure}
   \centering
    \includegraphics[width=\hsize]{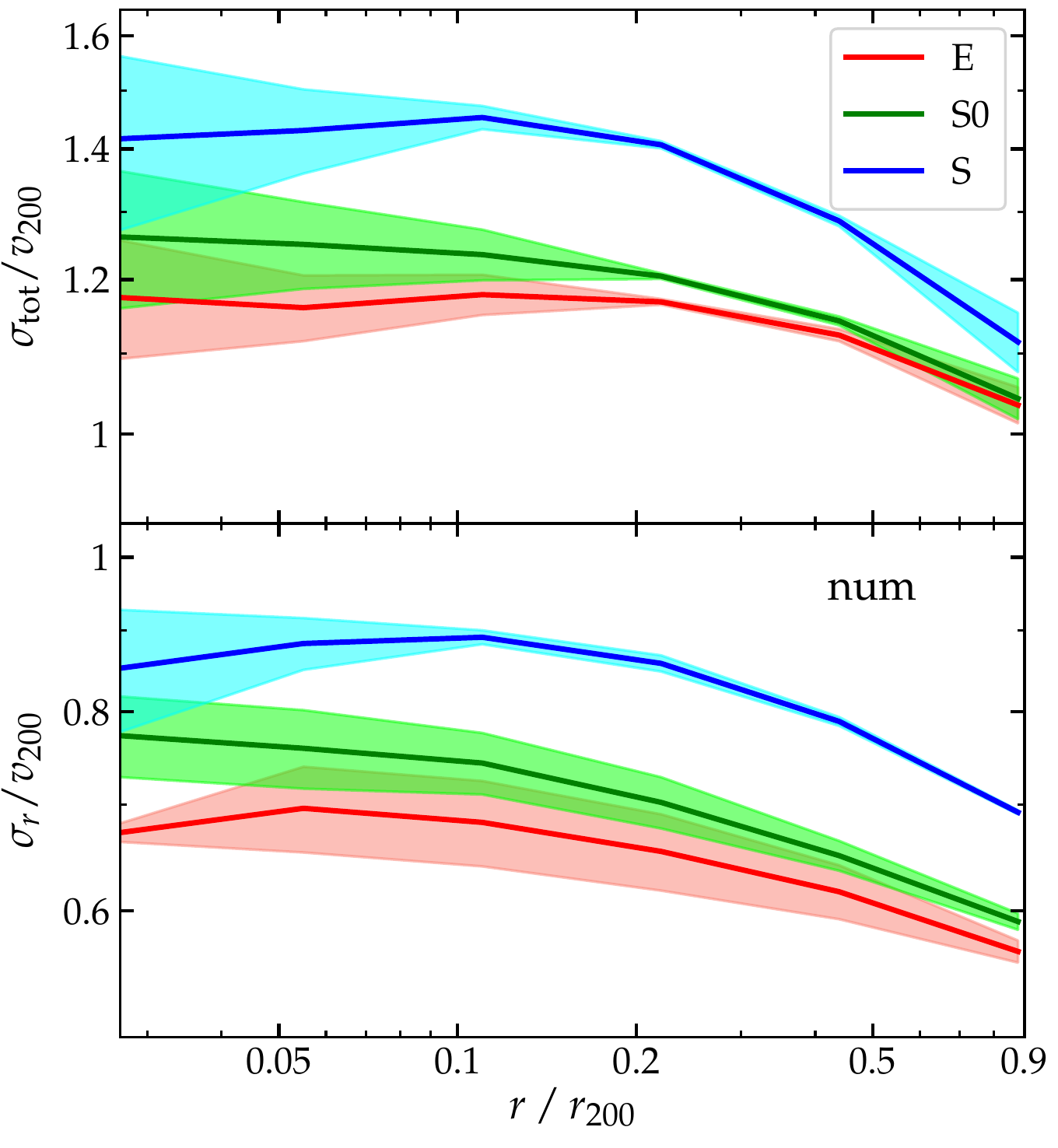}
   \caption{Same as Fig.~\ref{f:vdps} but for the \texttt{num} scaling.}
         \label{f:vdpsnum}
         \end{figure}
           \begin{figure}
   \centering
    \includegraphics[width=\hsize]{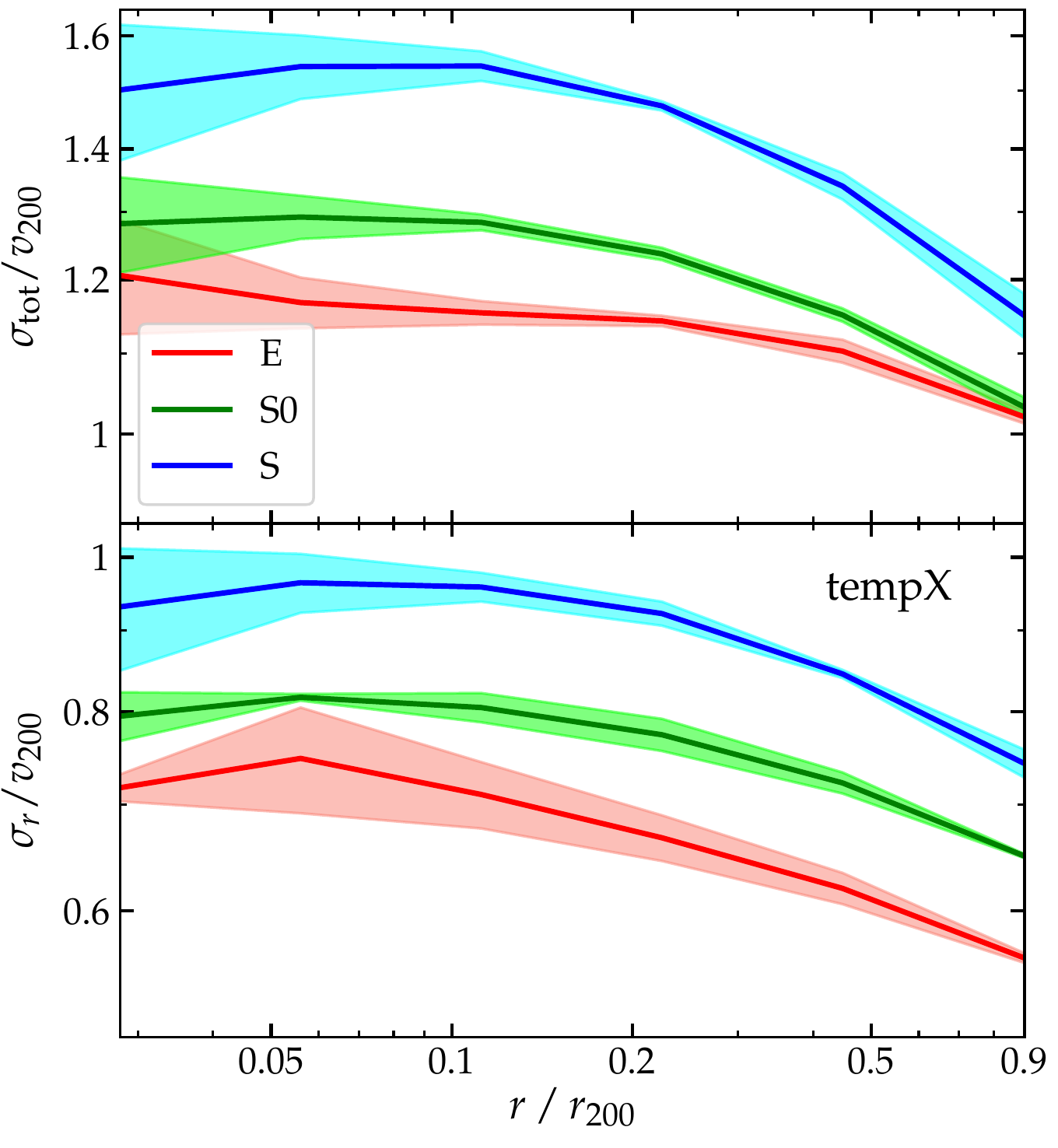}
   \caption{Same as Fig.~\ref{f:vdps} but for the \texttt{tempX} scaling. }
         \label{f:vdpstx}
         \end{figure}

\end{appendix}

\end{document}